# Magnetization Dynamics of Nanoscale Magnetic Materials: A Perspective

Anjan Barman[1,*], Sucheta Mondal[2], Sourav Sahoo[1], and Anulekha De[1]

[1]Department of Condensed Matter Physics and Material Sciences, S. N. Bose National Centre for Basic Sciences, Block JD, Sector III, Salt Lake, Kolkata 700 106, India

[2]Department of Electrical Engineering and Computer Sciences, University of California Berkeley, Berkeley, CA 94720-1770, USA

**\*Corresponding author email address:** abarman@bose.res.in

**Abstract:** Nanomagnets form the building blocks for a gamut of miniaturized energy-efficient devices including data storage, memory, wave-based computing, sensors and biomedical devices. They also offer a span of exotic phenomena and stern challenges. The rapid advancements of nanofabrication, characterization and numerical simulations during last two decades have made it possible to explore a plethora of science and technology related to nanomagnet dynamics. The progress in the magnetization dynamics of single nanomagnets and one- and two-dimensional arrays of nanostructures in the form of dots, antidots, nanoparticles, binary and bicomponent structures and patterned multilayers have been presented in details. Progress in unconventional and new structures like artificial spin ice and three-dimensional nanomagnets and spin textures like domain walls, vortex and skyrmions have been presented. Furthermore, a huge variety of new topics in the magnetization dynamics of magnetic nanostructures are rapidly emerging. An overview of the steadily evolving topics like spatio-temporal imaging of fast dynamics of nanostructures, dynamics of spin textures, artificial spin ice have been discussed. In addition, dynamics of contemporary and newly transpired magnetic architectures such as nanomagnet arrays with complex basis and symmetry, magnonic quasicrystals, fractals, defect structures, novel three-dimensional structures have been introduced. Effects of various spin-orbit coupling and ensuing spin textures as well as quantum hybrid systems comprising of magnon-photon, magnon-phonon and magnon-magnon coupling, antiferromagnetic nanostructures are rapidly growing and are expected to dominate this research field in the coming years. Finally, associated topics like nutation dynamics and nanomagnet antenna are briefly discussed. Despite showing a great progress, only a small fraction of nanomagnetism and its ancillary topics have been explored so far and huge efforts are envisaged in this evergrowing research area in the generations to come.

# I. INTRODUCTION

The quest of designing miniaturized and energy-efficient devices has led the scientific community to explore new paradigm of structuring and patterning materials into nanoscales. The journey started in the last century from extensive research on natural nano-composites to the patterning of artificial nanostructures[1,2]. The overriding goal has always been to defeat the excess energy consumption while enhancing storage capacity, operating speed and endurance of the device. For example, as the celebrated Moore's law is ending because of the physical limitations while increasing density of transistors on chip, it is imperative to search for suitable alternatives of charge-based semiconductor devices. Magnetic nanostructure has the potential to fulfill such demands. As the device performance relies on the nature of physical processes occurring within the nanomagnetic systems, researchers have invested relentless efforts for understanding these processes over the years. The development of state-of-the-art fabrication and characterization techniques has unraveled some exotic dynamical phenomena associated with various length and time scales. The magnetization dynamics refers to the dynamics of the population and phase of spins in an ensemble of particles. Some of the crucial parameters associated with spin dynamics are ultrafast demagnetization time, magnetization quenching, precessional frequency, relaxation time and magnetic damping, which play pivotal role in determining the efficiency of the nanostructure-based devices[3,4]. All these phenomena can be described in terms of magnetic interactions. The total free energy of a ferromagnetic system in presence of external magnetic field can be expressed as a sum of different competing energy terms[5]. These interactions are originated from the 'mixture' of atomic and macroscopic physics, which eventually govern the dynamics. The competition between these energies has prominent role in configuring magnetization distribution in the nanostructures which are completely different from their bulk states.

Magnetite ($Fe_3O_4$) nanoparticles that are involved in biomagnetic phenomenon and rock magnetism[6] and other ferromagnetic oxides found in ferrofluids and meteorites exemplify some of the naturally produced magnetic nanostructures. However, current research in nanomagnetism primarily deals with materials-by-design strategy. A fascinating approach is to develop engineered nanostructures, which are generally not encountered in nature. Engineered nanostructures can be of various types, such as, nanodots[7], nanoholes or antidots[8], nanostripes[9], nanowires[10], nanorings, nanoparticles[11], granular media, nanojunctions and multilayered nanostructures[12]. In the early days, the magnetization dynamics used to be focused in the single nanostructures and their monodimensional (1D) array. Later, the two-dimensional (2D) arrays were started to be fabricated and studied extensively. If an external perturbation is applied to a magnetic system, the exchange energy cost for a single spin reversal is reduced by spreading the disturbance over long wavelength. This propagation of exchange coupled spins forms the spin wave (SW) and the quanta of SW are known as magnons which can carry and process information. Recently, 'magnonics' has emerged as a sub-branch of 'nanomagnetism' which deals with the generation and manipulation of SWs in magnetic medium with periodically modulated magnetic properties offering periodic potentials to propagating magnons. Such artificial magnetic structures are termed as magnonic crystals (MCs)[13,14], which are the magnetic analogue of photonic and phononic crystals. By varying the geometrical parameters artificially, the nature of magnetic interaction and hence the SW dynamics can be modulated. More recently, hybridization of magnonics with other physical excitations has emerged new fields such as 'magnon-spintronics'[15], magphonics and magnon-polaritronics.

Current research focuses more on the application potential of magnetic nanostructures. Nanostructured multilayers (MLs) with magnetoresistive properties have become the bedrock of magnetic recording technology[12,16]. Nanomagnets are considered as the key elements for the various other applications, such as, bit patterned media[17], magnetic logic[18], data storage[19], spin-based transistors[20], reconfigurable waveguide for energy-efficient transmission of signal[21], spin-Hall nano-oscillators (SHNO)[22], spin-torque nano-oscillator (STNO)[23], neuromorphic and quantum computing[24] and biomedical devices. In the nanomagnetism family, MCs have shown potential for on-chip microwave communication and processing over the electromagnetic wave-based devices due to their inherent non-reciprocity, energy efficiency and nanoscale wavelength for GHz to sub-THz frequency range. Plethora of studies on the quasistatic and dynamic properties of 1D and 2D MCs have been carried out to explore their fundamental SW properties besides their promising applications in SW logic, resonator, filter, phase shifter, splitter, directional coupler and many other magnonic devices[2,14,15,25]. The extension towards the third dimension (3D) of magnetic nanostructures may give rise to many complex magnetic configurations with unprecedented properties[26]. In spite of such advancements, nanomagnetism still faces many stern challenges. Harnessing the full potential of the magnetic nanostructure needs further experimental and theoretical groundworks in the coming years.

The ultrafast magnetization dynamics of magnetic nanostructures involve complex and nontrivial physics. In this perspective article, we have deciphered some of these complexities by introducing clear classification of nanostructured materials in different length scales and associated magnetization dynamics in characteristic time scales. Detailed discussion about the static and dynamic properties of the nanostructures having periodicity in different dimensions, is presented based on the available literature. In addition, multilayered nanostructures and their application potentials have been briefly described. We have introduced some of the recent developments in the area of 3D nanostructures. We have also shared our perspective on the emergent phenomena like topological magnetic textures, strong coupling, antiferromagnetic nanomagnets, ultra-high resolution spatiotemporal imaging of SWs etc. We have finally illustrated the scope in these fields with the hope that further research in these directions will help to unlock new computational paradigms with ferromagnetic nanostructures.

## II. OUTLINE OF THE PROBLEM

Fabrication of the nanostructures with high quality lateral features, surface and interface with dimensions down to almost atomic range involves massive effort by the experts in the field of nanoscience and technology. Several electrical, optical, electro-optical and atomic interaction-based characterization techniques have been developed relentlessly to probe their magnetic properties. One of the major goals to this end have been to achieve unprecedented spatiotemporal resolution. The quantitative and qualitative characterization of SW dynamics of the nanostructures can now be easily performed in space, time, frequency, wavevector and phase domains. Various micromagnetic and atomistic simulators along with theoretical approaches have been developed to underpin the intriguing physics of magnetic nanostructures. In this section we have articulated an overview of the background of this field.

### A. Importance of length and time scales

The spin configurations undergo enormous variation while sculpting down the magnetic material from micro to nanoscale. The energy surfaces and hence the ensuing static and dynamical properties modify drastically. For example, micron-sized soft-magnetic disk accommodates magnetic vortex state with in-plane spin configuration and an out-of-plane core. When the size is gradually reduced to nanoscale the system undergoes a phase transition via quasi-single domain to single domain structure. Magnetization dynamics can be classified as follows based on the characteristic time scales: the fastest process is the fundamental exchange interaction occurring within about 10 fs. The spin-orbit coupling (SOC) and related phenomena also occur in the fs temporal scale. The laser-induced ultrafast demagnetization takes place within from sub-hundred fs to sub-ps range. The fast remagnetization covers the time span of few ps which is followed by a slow remagnetization, precession, damping and SW dynamics over few hundred of ps to ns time scale. The relatively slower processes are vortex core gyration, core switching (hundreds of ps of several ns) and domain wall (DW) motion (few ns to µs). A pictorial description of various phenomena in different time and length scales is presented in Fig. 1 (a) and (b).

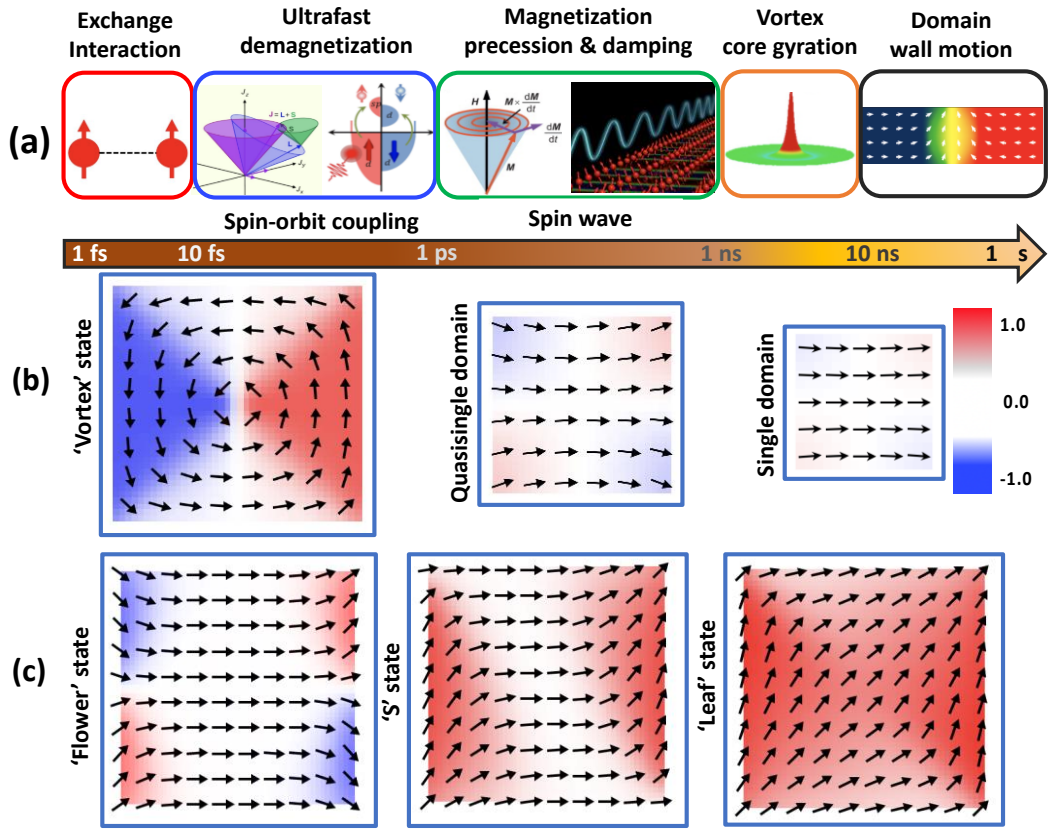

FIG. 1. (a) Schematic representation of time scale of magnetization dynamics. Representative spin configurations of nanomagnet with varying (b) size and (c) applied magnetic field orientation. The corresponding color bar is shown at the right-hand side of the figure.

**B. Overview of sample fabrication techniques**

The state-of-the art fabrication techniques developed over last few decades have brought a revolution to the accessible length scales unfurled from micro- to nanoscale. Furthermore, the continuous hunt for miniaturized devices has urged the scientists to push the boundary beyond atomic limits. High quality samples can be prepared by using both top-down and bottom-up approaches.

**Top-down approach**

The top-down approach relies on lateral patterning of bulk materials by subtractive or additive methods. Some of the lithography methods which follow this approach are photolithography, electron-beam lithography, deep ultraviolet lithography, ion beam lithography, scanning lithography, soft lithography, nanosphere lithography, scanning probe lithography, colloidal lithography etc.[27-29]. There are several other methods under this category, such as, shadow masking, laser or ion-beam irradiation, nanocontact printing, ion implantation, laser machining, deposition and diffusion etc. This approach is reliable to fabricate range of structures, *viz.* nanodot array, antidot array, nanostripes, quasicrystals, patterned magnetic thin films and MLs with high repeatability but low yield.

**Bottom-up approach**

The bottom-up approach relies on the chemical synthesis and mesoscopic pattern formation. Thus nanoparticles, nanowires, micro-organisms are the nanostructures that can be synthesized using this approach by using suitable templates, such as, track-etched polymers, anodic alumina, di-block copolymer membranes[30,31] etc. Several methods which follow this approach are, plasma arcing, chemical vapor deposition, molecular beam epitaxy, sol gel synthesis, molecular self-assembly etc. More recently, genetically or mechanically modified magnetosomes from magnetotactic bacteria have been used to grow well controlled structures for biologically encoded magnonics[32]. However, it is worth mentioning that a much-improved controllability will be required to make this process competitive for device fabrication.

In the recent past, one of the major hurdles was to fabricate high quality 3D structures. Two-photon lithography (TPL) combined with the electrodeposition technique[33], laser micromachining, 3D printing, advanced chemical synthesis etc. have emerged as powerful techniques to cross this hurdle.

**C. Characterization of magnetic properties**

Magnetic properties of these nanostructures can be characterized by using several techniques which conduct beyond conventional magnetometry the imaging of the spin configuration, and measurement of quasistatic and fast dynamic properties. Besides, there are various numerical tools available to simulate these properties.

**Imaging**

Intricate properties such as spin textures and stray field distributions in magnetic nanostructures can be probed using magnetic force microscopy (MFM), magneto-optical Kerr effect (MOKE) microscopy[34,35], Lorentz transmission electron microscopy (LTEM) etc. Spatial resolution of MFM is bounded by the tip curvature and by optical diffraction limit for MOKE microscopy. Thus, it was necessary to implement high resolution techniques like LTEM, spin polarized low-energy electron microscopy (SPLEEM), photo-emission electron microscopy (PEEM), spin-polarized scanning tunneling microscopy (SPSTM), scanning electron microscopy with polarization analysis (SEMPA), and ballistic electron magnetic microscopy (BEMM) to successfully image the spin distribution combined with topographic characteristics[36]. Electron holography technique also provides amplitude and phase information of the magnetic nanostructures with resolution down to 2 nm.

**Characterization of quasistatic magnetic properties**

The static magnetic properties of nanostructures, such as magnetic moment, coercivity, saturation field and Curie temperature can be characterized by using superconducting quantum interference device (SQUID) magnetometry with high sensitivity, whereas vibrating sample magnetometry (VSM) is suitable for characterizing relatively thicker films, bulk materials and large assembly of magnetic nanoparticles. MOKE in different geometry, such as, longitudinal, transverse and polar, is very useful technique to characterize the quasistatic magnetic properties in a local and noninvasive manner. Several magneto-resistive methods and Hall effects are used to probe anisotropic magnetoresistance (AMR), spin Hall magnetoresistance (SMR), anomalous Hall effect (AHE) etc.

**Characterization of dynamic magnetic properties**

Many state-of-the art techniques have been developed to excite and probe the magnetization dynamics in magnetic nanostructures occurring over microsecond to femtosecond regimes. Among those, time-resolved MOKE (TRMOKE), ferromagnetic resonance (FMR) and Brillouin light scattering (BLS) technique are three most well-used techniques to characterize the magnetization dynamics in time, frequency, wave-vector, space and phase domains[4,37]. Figure 2 shows the schematic diagram of some of these microscopy techniques. TRMOKE technique can be categorized into field-pumped and optically-pumped approaches. The temporal resolution is determined by either the field pulse width (in field-pumped technique) or the optical pulse width (in case of optically-pumped technique) and together they can resolve ultrafast demagnetization, remagnetization, magnetization precession, damping, SW confinements, nonlinear optical effects, optically induced spin transfer (OISTR), strong coupling effects as well as carrier and phonon dynamics. This offers a local measurement and can probe the magnetization dynamics by avoiding linewidth broadening due to spatial nonuniformities. Modal composition and damping of individual modes can also be reliably characterized with this technique. Time-resolved scanning Kerr microscopy (TRSKM)[38] facilitates the imaging of temporal evolution of spatial magnetization distribution by fixing the time delay and using scanning MOKE microscopy at various delays. However, the diffraction-limited spatial resolution is only few hundred nm, which can go down to sub-100 nm by using near-field MOKE[39]. In the frequency domain, ranging from MHz to tens of GHz, broadband FMR is a very useful technique to globally excite the sample by using absorption from the external source in the frequency spectrum. It has other variants, such as, conventional resonant cavity FMR, spatially resolved FMR, spin-torque FMR (ST-FMR)[40-42] etc. However, delicate micro-fabrication of waveguide structures on the sample makes this technique more cumbersome. BLS technique is primarily advantageous due to its wave-vector sensitivity which allows the measurement of SW dispersion and magnonic band structure of various magnonic crystals[43]. Magnon band gaps and group velocity can also be reliably studied. Another advantage of BLS is that it is a noninvasive technique where the SW dynamics can be excited using thermal energy at room temperature and hence no external excitation source and synchronization of the probe with the external source is required. However, by launching rf-current, spin torque or other external stimuli at specific resonant frequencies, the SW propagation for that frequency can be spatially mapped in micro-BLS technique. It can also have variants like time-resolved and phase-resolved BLS, making it one of the most versatile techniques for SW measurement. All these bench-top techniques are extremely sensitive and require extreme care and regular maintenance. Apart from these popular tools, pulse inductive microwave magnetometer (PIMM) is an oscilloscope-based time-domain detection tool to measure the dynamics with tens of ps resolution. Another recently evolved imaging technique is magnetic resonance force microscopy (MRFM), bridging the gap between magnetic resonance imaging (MRI) and scanning probe microscopy (SPM). Attosecond spectroscopy, higher-harmonic generation, neutron scattering, X-ray microscopy based on X-ray magnetic circular- and linear-dichroism (XMCD & XMLD) processes are high-end facilities which can probe the static as well dynamic magnetic response from the ferromagnetic, ferrimagnetic and antiferromagnetic thin films and nanostructures.

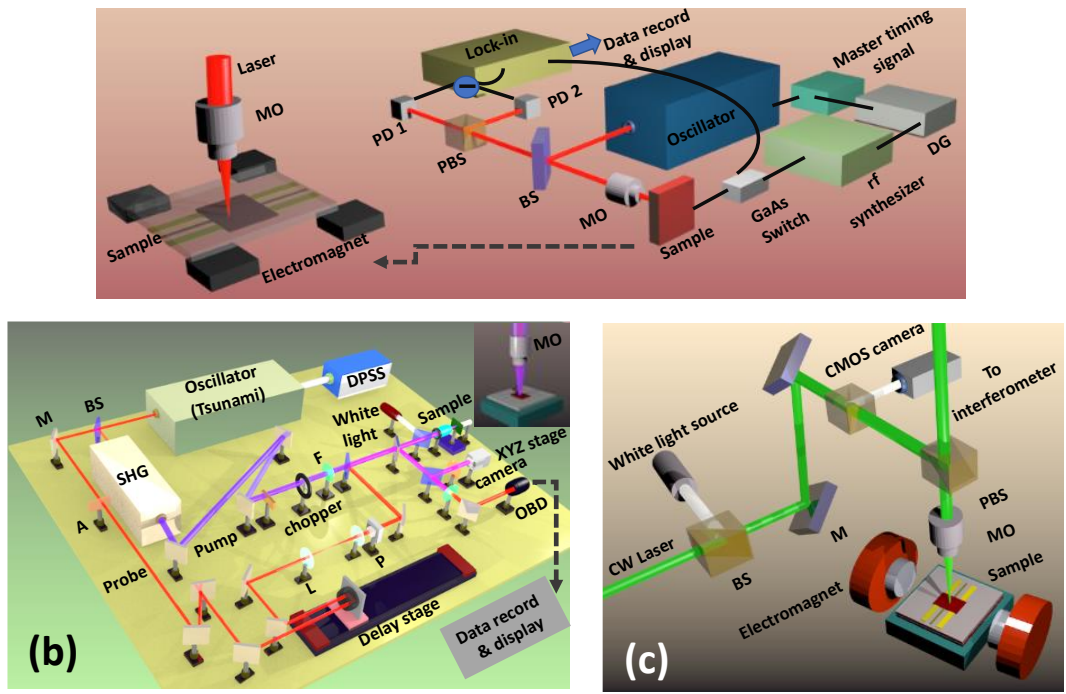

FIG. 2. Schematic diagram of (a) the spatially resolved ferromagnetic resonance (FMR) technique, (b) the time-resolved magneto-optical Kerr effect (TRMOKE) microscopy, and (b) the Brillouin light scattering (BLS) microscopy. The notations for the components are given as: BS = beam splitter, PBS = polarized beam splitter, M = mirror, A = attenuator, L = lens, OBD = optical bridge detector, F = filter, P = polarizer, PD = photodiode, DG = delay generator, MO = microscope objective, SHG = second harmonic generator, DPSS = diode-pumped solid state laser.

**Numerical calculations**

Micromagnetic and atomistic simulations, alongside density functional theory and Monte Carlo methods, have played pivotal roles in predicting a wide variety of magnetic phenomena in confined magnetic systems over the years. A series of micromagnetic simulators have been developed which are extremely useful to simulate the magnetization dynamics covering a broad timescale. Object oriented micromagnetic framework (OOMMF), LLG Micromagnetic Simulator, MicroMagus, magpar, Nmag, mumax$^3$, MagOasis, Magnum, Fidimag, Boris, etc. are all popular softwares which are based on finite difference or finite element method. Atomistic simulators such as, Vampire, UppASD, Spirit, Fidimag etc. have been developed to simulate magnetic materials with atomic resolution from Angstroms to several micrometers. Micromagnetic simulations cannot precisely capture the microscopic origin of complex physical effects such as exchange bias, spin-orbit effects, spin transport, heat assisted magnetic effects, ultrafast demagnetization etc. Atomistic simulations can bridge the gap between electronic structure and micromagnetic method by treating the material at its natural atomic length scale. Numerous theoretical models and packages are available to reliably calculate the electronic band structures and electronic properties of magnetic materials. To this end plane-wave method (PWM) based numerical calculations have become popular for calculating magnonic band structures in 1D, 2D and 3D MCs.

**D. Intrinsic dynamics of isolated nanomagnet**

With spatial confinement, domain formation should be completely suppressed at the nanoscale according to Brown's fundamental theorem. An individual nanomagnet should thus behave as a giant spin comprising of numerous isotropic spins compactly residing together. However, any non-ellipsoidal nanomagnet exhibits anisotropy arising from its geometric configuration other than magnetocrystalline and magnetostrictive anisotropies. Cowburn *et al.* in 1998, reported a strong configurational anisotropy in a single square-shaped supermalloy nanomagnet having sub-micon size by using MOKE technique[44]. Presence of shape-dependent four-fold and weak eight-fold anisotropy were evidenced. Confinement from micrometer through sub-micrometer to nanometer scale imposes several phase transitions. In 1999, another work by Cowburn *et al.* showed that magnetization reversal mechanism undergoes from a squeezed hysteresis loop to a square loop. This indicates the presence of two different phases in circular nanomagnets: vortex and single domain states[45] according to their aspect ratio (thickness/radius). Additionally, the spin configuration can vary with relative orientation between applied magnetic field and anisotropic field which is shown for a square-shaped nanomagnet in Fig. 1 (c).

In 2004 Demidov *et al*. performed a pioneering experiment with square-shaped individual Permalloy (Py hereafter) magnet with sub-micrometer size fabricated on top of a CoFe microstructure separated by copper spacer[46]. They demonstrated this individual magnet with typical memory-element like geometry, as a genuine source of microwave radiation. The micro-focused BLS measurement showed radiation of SWs into the surrounding magnetic film occurring at discrete frequencies corresponding to the frequencies of quantized modes of the element at around 10 to 12 GHz. It is pertinent to mention here that very rapidly the researchers realized the importance of high-aspect ratio nanomagnets to be embedded in the microwave devices and the focus immediately shifted to nanoscale elements. The magnetic-field dispersion of SW frequencies obtained in BLS measurements supported by micromagnetic analysis revealed that a nanomagnet can accommodate a stable center mode (CM), an edge mode (EM), and confined standing wave modes of Damon-Eshbach (DE), backward volume (BV) and hybrid nature[47].

The ambition of probing the temporal variation of magnetization in single nanomagnets well below the diffraction limit led to the development of cavity-enhanced MOKE (CEMOKE) technique[48]. In 2006, Barman *et al.* reported the ps dynamics of Ni nanomagnet with different size by exploiting this highly sensitive CEMOKE technique[49]. A significant speed up in the magentization dynamics from sub-GHz to GHz frequency range was observed as the cylindrical magnet enters from multidomain- to single-domain state. Also, with the decrease in diameter of these cylindrical nanomagnet the damping of the coherent precessional mode decreased sharply to finally settle down to its intrinsic value in the single domain regime. The bias field dependent precessional frequency confirmed an extrinsic contribution to damping in micrometer sized nanomagnets, where a lower frequency mode overlapped with the fundamental mode with decreasing field, causing significant dephasing[50]. However, such extrinsic contribution was not observed for the nanomagnets leading towards a field-independent damping. During the same time Laraoui *et al*. reported the ultrafast demagnetization and precession[51] and relaxation[52] of $CoPt_3$ sub-micrometer sized single dots using an all-optical TRMOKE system having femtosecond temporal resolution and high spatial resolution (~300 nm) obtained with a reflective confocal Kerr microscope. In particular the fast relaxation time of few ps, where electrons and spins exchange energy with the lattice and the slow relaxation time of hundreds of ps, when electrons and the lattice exchange energy with the environment were reported from single nanomagnets. In 2008, Liu *et al*. measured the time-resolved magnetization dynamics of an individual Py disks of 160 nm diameter using time-resolved Kerr microscopy. By sweeping the bias magnetic field the internal spin configuration underwent transition between the vortex and quasisingle domain states, leading to distinct hysteresis behavior of fundamental mode frequency as a function of the in-plane bias field, and the critical fields for triggering the vortex annihilation and nucleation processes have been determined in this study[53]. The year 2011 witnessed some very important developments in single nanomagnet dynamics. Rana *et al*. measured the time-resolved dynamics of isolated 50-nm-wide Py dot showing a dominant edge mode without having any trace of centre mode[54]. They also observed quadrupolar interaction between the nanodots with increased areal density and a dynamic dephasing for the enhancement of damping in agreement with previous theoretical work from the same group[55]. On the other hand, Liu *et al*. detected high-frequency dynamics of a single 150 nm wide nanomagnet from a lower frequency background of 500 nm wide nanomagnets by placing them in the same array. They claimed that the optical diffraction limit could be beaten since the characteristic behavior of the studied nanomagnet is sufficiently different from its neighbors[56]. Naletov *et al*. used MRFM to study radial and azimuthal eigenmodes in a Py/Cu/Py spin-valve-like nanopillar by applying spatially uniform rf field or rf current flowing through the nanopillar and found a selection rule for exciting different modes by adjusting the excitation geometry[57]. Keatley *et al*. demonstrated controlled suppression of EM in an individual nanomagnet by excitation of larger amplitude coherent precession of CM. This is necessary for nanoscale spin transfer torque (STT) oscillators and bi-stable switching devices where more uniform spin dynamics is desirable[58].

The focus steadily shifted towards the using unconventional external stimuli for excitation of single nanomagnet dynamics. It has been known that nanomagnets driven by spin-polarized current can exhibit high frequency magnetization dynamics and can act as a microwave resonator[59]. Further, the antidamping torque from pure spin current can set the magnetization precession to auto-oscillation by suppressing its intrinsic damping. Investigation showed that spin-orbit torque (SOT) generated from adjacent heavy metal layer to a ferromagnet can switch the nanomagnet in a deterministic way, which can be used to construct a field free clocking and nanomagnetic logic analogous to current CMOS technology[22]. All-optical helicity dependent switching in single nanomagnet reveals intriguing physics which can lead to direct and fast data writing. To this end, the observation of smaller nanoelements settling to their final magnetization states faster (~2 ps) after switching than larger elements ushers new hope[60]. The faster switching speed is attributed to the electron-lattice and spin-lattice interactions with higher spin temperatures for smaller nanoelements[61]. Strong coupling of magnon with photon, phonon or other magnon can lead to hybrid systems for quantum transduction. To this end direct observation of strongly coupled magnon–phonon dynamics via the orientation of the applied magnetic field was an important development[62]. Besides, strain-induced switching has found to establish a successful Boolean operation in recent past[63]. Recently, hybrid magneto-dynamical modes have been observed in a single quasi-elliptical magnetostrictive Co nanomagnet deposited on PMN-PT substrate using TRMOKE microscope (see Fig. 3)[64]. The intense electric field of a pump laser induced periodic strain in the substrate and generated surface acoustic waves (SAWs), giving rise to an oscillation in the nanomagnet's magnetization via Villari effect. Hybrid modes, *viz.* mixed mode with azimuthal quantization at higher frequency and tilted EM in the lower frequencies, were generated due to the dynamical mixing of the periodic strain induced magnetization oscillation and laser induced precessional motion. This could enable development of strain actuated magnetoelastic nano-oscillator and magnonic logic circuits.

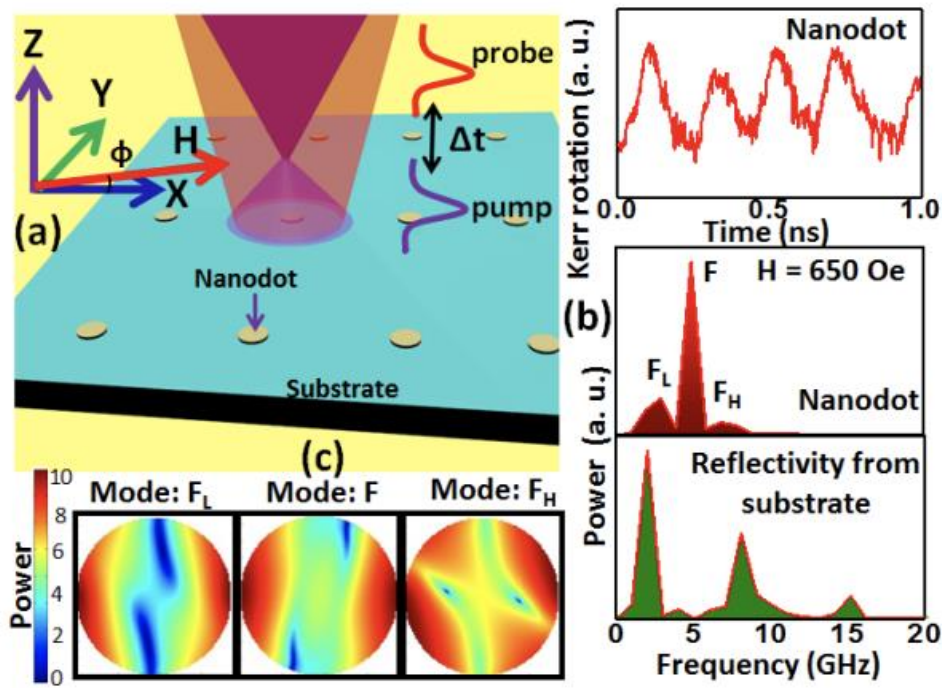

FIG. 3. (a) Schematic diagram of the experimental geometry. (b) Kerr rotation data and corresponding frequency spectra obtained from Co nanodot for bias magnetic field $H$ = 650 Oe. The peaks in the latter spectrum correspond to the frequencies of the SAWs excited in the PMN-PT substrate. (c) Simulated power profile of the SW modes in the nanomagnet. Adapted with permission from S. Mondal *et al*., ACS Appl. Mater. Interfaces **10**, 43970 (2018). Copyright 2018 American Chemical Society [64].

**E. Dynamics of 1D arrays of magnetic nanostructures**

1D ferromagnetic nanostructures in the form of ordered arrays of ferromagnetic nanostripes or nanowires have attracted considerable interest, where the magnetic shape anisotropy plays a significant role, leading to the bi-stable magnetic states. These nanowires are important candidates for racetrack memory, interconnects, magnonic waveguides and nanochannels[65,66]. 1D magnonic waveguides have been realized in the form of nanostripes with periodic modulation of geometric (e.g. thickness or width of nanowires) or magnetic (such as the magnetic anisotropy, saturation magnetization, strength and orientation of magnetic field) parameters[67,68] which are important components of integrated magnonic nanocircuits. Experimental investigations have demonstrated that one can efficiently channel, split, and manipulate propagating SWs, in magnonic waveguides[69]. A key requirement of magnonic nanocrircut is to have SWs turning a sharp bend or a corner without significant dissipation, which was successfully demonstrated in a 'S' shaped bend in a Py waveguide[70]. Numerical simulations have shown the opening of magnonic band gap (MBG) by introducing a row of antidots in a magnonic waveguide (MAW)[71], which makes them a promising candidate for magnonic nanocircuit with integrated filter. Efficient engineering of magnonic band structure has been achieved by mirror symmetry breaking in a MAW and subsequent application of an external magnetic field[72]. All-optical TRMOKE study in Py nanostripes showed a stark variation in the frequency, anisotropy and spatial nature of SWs depending on the stripe width and the orientation of the magnetic field[73]. Single-crystal Ni nanowires with high aspect ratio revealed that standing and uniform SW modes can merge to a form a single uniform SW mode by tuning magnetic field[74]. Nonlinear effect of propagating SWs has also been observed in ferromagnetic microstripes[75]. Width modulated stripes have been claimed to have rich potential for magnonic band tunability simply by varying the modulation parameters[76] and only few experimental studies exist to that end[77]. A recent study showed that the MBG in an asymmetric saw-tooth-shaped (ASW) ferromagnetic array forming a pseudo-1D MC can be easily reconfigured by changing the orientation of the applied magnetic field[78].

**F. Dynamics of magnetic nanoparticles**

Chemical synthesis of magnetic nanoparticles (MNPs) are relatively easier compared to nanofabrication based on lithography. Their dimensions can reach close to the atomic scale and they are susceptible to external magnetic field, and hence, they find applications in multidisciplinary fields, such as, nanoelectronics, magnetic data-storage, sensors, contrast enhancement in magnetic imaging, drug delivery and other bio-medicinal applications[79,80]. Magnetization reversal of MNPs have been studied extensively, and it was shown that the 2D Stoner-Wohlfarth model is too simple and higher order anisotropy terms are essential to understand their magnetization reversal dynamics both in classical and quantum limits[81]. Further, the issue of high switching field of MNPs with smaller size have been resolved by simultaneous application of a small dc field with a rf field in a 20-nm-diameter Co particle using micro-SQUID measurement[82]. The magnetization reversal mechanisms for assembly of MNPs mediated by the exchange and dipolar interactions have been studied in detail[11,83].

The first measurement of ultrafast spin dynamics of MNPs was reported by Buchanan *et al.*, who studied the magnetic field pulse-induced precessional dynamics of Fe nanocrystals of 25 nm size embedded in $SiO_2$ matrix[84]. Observation of high resonance frequency, strong effective damping, and electrically insulating character of the samples were considered favourable for application in ultrafast magnetic sensors. Andrade *et al*. measured the optical pulse induced coherent precessional dynamics in superparamagnetic Co particles with size down to 2.5 nm in $Al_2O_3$ matrix. For the smallest particles the precession was found to be critically damped, preventing their coherent magnetization reversal. A complete gyroscopic pathway during magnetization reversal of small superparamagnets was therefore considered unlikely[85]. The time-resolved ultrafast demagnetization and subsequent relaxation processes of colloidal ferrites showed a strong size and magnetic ordering dependence of the amount of magnetization quenching and recovery but no such variation in the demagnetization time. The partial re-establishment of ferrimagnetic ordering before electronic relaxation was correlated to the faster recovery (~2 ps) while the slow recovery was correlated to the electronic relaxation[86]. The same authors investigated the time-resolved magnetization dynamics of $Fe_3O_4$ nanocrystals with size down to 5 nm to understand the relative efficiency of the spin-lattice relaxation on the surface of the nanocrystals with respect to its interior[87]. In Co(core)-Pt(shell) nanoparticles with 5 nm core diameter and 1.5 nm shell width, the effect of hard laser annealing was found to cause diffusion of Pt into the Co core, resulting in the formation of CoPt alloy and an increase the magnetocrystalline anisotropy. The latter caused a femtosecond laser induced GHz frequency magnetization precession in CoPt NPs which was absent in their superparamagnetic counterparts[88]. Rana *et al*. reported a cluster configuration independent ultrafast demagnetization in Ni NPs with chain, bundle, dendrite and random assembly[89]. However, the fast- and slow-recovery times and precessional dynamics were found to be strongly dependent on the agglomeration geometry due to their internal distribution and interaction with the environment. The precession frequencies decreased with the ordering in the agglomeration geometry most likely due to the decrease in the shape anisotropy. More recently, measurement of spatially resolved demagnetization inside FePt nanoparticles using time-resolved coherent X-ray scattering showed inhomogeneous demagnetization within the nanoparticles, which occurred more rapidly at the boundary of the nanoparticle. It further showed formation of the shell region with reduced magnetization driven by a superdiffusive spin flux and its inward propagation at a supermagnonic velocity[90]. Various theoretical models of nanoparticle dynamics and computational nonequilibrium models[80] exist to underpin the intriguing physics of these system and it is still an open field of research.

**G. Dynamics of 2D arrays of nanostructures**

Patterning of magnetic thin films in 2D down to the nanoscale can lead to connected and disconnected structures such as magnetic nanodots, antidots and their composites with the recent advancement of fabrication technologies. Based on their structural features they can be broadly classified as: nanodot, antidot, nanoring, bicomponent and binary nanostructures. In this subsection, we will briefly discuss their high frequency dynamics.

**Magnetic dots**

The early theoretical studies of high-frequency response of ferromagnetic nanodot arrays commenced in the late 1990s, where the effects of inter-dot dipolar coupling on the dynamics by varying array geometry and magnetic field orientation have been studied[91,92]. The initial experimental studies of FMR spectra of periodic nanodot arrays showed decomposition of single resonance peak into multimodal oscillations, whose position strongly depends on the orientation of the external magnetic field and the interparticle interaction[93]. This was followed by slow upsurge of experiments in this field, *viz.* BLS studies of cylindrical Py nanodots showing two classes of modes, namely higher-frequency DE-like and lower-frequency BV-like SW branches[94], spin excitation in similar nanodots over wide magnetic field range covering uniformly magnetized and vortex state[95] and FMR study of perpendicularly magnetized nanodots showing a large number of modes independent of interdot separation stemming from dipole-exchange interaction[96]. This was followed by the study of time-resolved magnetization dynamics in square shaped $Co_{80}Fe_{20}/Ni_{88}Fe_{12}$ bilayer nanodot arrays with varying size down to 64 nm showing a non-monotonic variation of the precession frequency with dot size owing to a crossover from CM to EM domination below a dot diameter of 220 nm[97]. This was followed by important observations like dynamical configurational anisotropy[98] and SW modes in nonellipsoidal elements with nonuniform ground states[99]. In 2009 Shaw *et al*. used frequency-resolved MOKE experiment in Py nanoelements having diameter between 50 – 200 nm to observe that intrinsic Gilbert damping parameter is generally unaffected by the nanopatterning process despite a large linewidth dependence on the size of the nanomagnets. The linewidth of the CM and EMs also found to differ considerably, most likely due to the sensitivity of the EM to small variations and imperfection of the shape and edge materials[100]. In the next decade the focus shifted towards the observation of collective magnonic modes in strongly coupled arrays of nanoelements. In 2010, collective magnonic modes were detected in an array of closely spaced elements where the array appears as tailored magnonic metamaterials to spin and electromagnetic waves with a wavelength well beyond the period of the array[101]. Anisotropic dynamic coupling for propagating collective modes were observed in 2D arrays of square elements[102]. Further study in square shaped Py nanodots arranged with varying areal density showed a transition from a single uniform collective mode at very high areal density through weakly collective dynamics at intermediate areal density to completely isolated dynamics of the individual nanodots at very small areal density[103]. In the same year collective SW excitations in the form of Bloch waves propagating through chains of dipolar coupled nanodots characterized by magnonic energy bands were reported[104]. Bondarenko *et al.* showed collective modes for a ferromagnetic dot array with perpendicular magnetization[105]. Effects of lattice symmetry started to be explored when magneto-dynamical response of large-area close-packed arrays of circular dots on hexagonal lattice fabricated by nanosphere lithography was reported. Saha *et al.* reported a comprehensive study of the effects of varying lattice symmetry where a stark variation of collective SW modes was observed when the lattice symmetry of circular-shaped Py nanodot arrays were varied from square to octagonal through rectangular, hexagonal and honeycomb lattices[106]. Effects of dot shape were also explored where a cross-shaped nanoelement showed strong anisotropic behaviour[107] and varying shape like elliptical, half-elliptical, rectangular, triangular and diamond-shaped elements showed strong variation in SW spectra due to internal field profile and the ensuing mode quantization[108]. Figure 4 shows the effects of dot shape on the bias-field dependent SW frequencies for 2D nanodot arrays with the above-mentioned shapes. The values of the effective demagnetization factors have been extracted from the collective modes as shown in the figure caption. The cross-shaped element continues to show great prospects with observations like mode softening, mode crossover, mode splitting and merging of SW frequency branches with the bias field strength and orientation[109] and more recently a strong magnon-magnon coupling and nonlinear FMR behaviour[110].

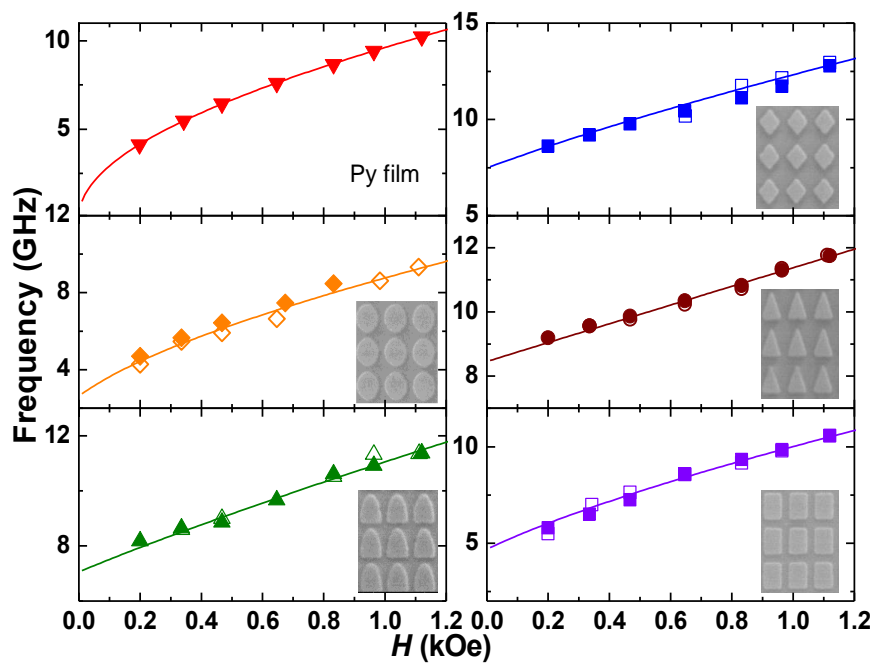

FIG. 4. Bias magnetic field dependent frequency of the uniform precessional mode of 20-nm-thick $Ni_{80}Fe_{20}$ thin film and nanodots arrays with varying dot shapes having lateral dimensions of 200 nm × 250 nm of individual dots and edge-to-edge interdot separation of 75 nm. The filled symbols show experimental data obtained from TRMOKE experiment and open symbols are micromagnetic simulation results. The solid line is a fit with Kittel formula: $f = \frac{\gamma}{2\pi}\sqrt{(H + (N_z - N_x)4\pi M_s)(H + (N_z - N_y)4\pi M_s)}$, where $N_x$, $N_y$ and $N_z$ ($N_x+N_y+N_z = 1$) are the demagnetizing factors for the nanodots. From the fits we obtained ($N_x$, $N_y$, $N_z$) as (0.103, 0.193, 0.704) for elliptical dots, (0.011, 0.079, 0.910) for diamond dots, (0.137, 0.337, 0.526) for triangle dots and (0.027, 0.047, 0.926) for rectangular dots. However, the obtained demagnetizing factors contain contributions from both the individual elements and the stray field mediated inter-element interactions. The SEM images of the nanodots are shown in the insets.

**Magnetic antidots**

Magnetic antidot lattices (ADLs), *i.e.* periodically perforated ferromagnetic thin films are considered to be a strong candidate for designing reconfigurable MCs. The ADLs have some unique advantages over the isolated nanomagnet arrays due to the absence of any small isolated magnetic entity. Here, the entire film remains exchange coupled and hence offers higher SW propagation velocity and longer propagation distance for the SWs as opposed to isolated nanomagnets. The ADLs can also be described as a mess of connected networks which do not suffer from the superparamagnetic lower limit as opposed to the isolated nanomagnets. Extensive experimental and numerical investigations on the dynamics of standing and propagating SWs in magnetic antidots have led towards important findings.

Martyanov et al.[111] reported the first experimental study of the magnetization dynamics of Co antidot arrays by FMR, which showed evidence that characteristic inhomogeneities in the magnetization distribution around the antidots give rise to the changes of the resonance modes with the in-plane direction of the magnetization. This was immediately followed by study of SW localization between nearest and next nearest holes of ADL[8]. This was followed by observation of magnonic normal mode[112], Bloch-wave mode and an unusual bias field independent mode[113] in ADL. Anisotropic propagation, damping and velocities of SWs with bias field orientation[114] was an important development, besides the tunability of transmission coefficient of SW by the orientation of external magnetic field claiming a tunable metamaterial response[115]. Some key phenomena of SW dispersion in magnetic ADL appeared in a flurry, e.g. Bragg diffraction of SW from ADL and ensuing MBG[116], high-symmetry magnonic modes in perpendicularly magnetized ADL[117], and complete MBG for magnetostatic forward volume waves in 2D ADL[118]. Subsequently external and internal control of SW modes in ADL started by varying lattice constant[119], antidot shape[120], lattice symmetry[121], base material[122,123], bias field strength and orientation[114,124]. Furthermore, several interesting phenomena such as mode conversion[125], mode crossover and mode hopping[126], mode softening[127], as well as the formation of magnonic mini band[128] have been reported.

Defects play important roles in MCs. It can either be inadvertent defect originating from nanofabrication or tailored defect which can be used to our advantage for further tunability of magnonic bands and creation of defect states. Numerical studies showed that the magnonic spectra of hexagonal array of antidot is quite robust to random defects[129]. Introduction of a line defect on the other hand showed elevated frequency of the fundamental mode due to the increase in internal field in AD-free region and generation of a new extended mode with wider profile[130]. Extensive study of defects showed softening of the EM and localized modes accompanied by a possible amplification of the extended modes at quasi-saturation fields and to a local alteration of SW mode profiles. At low fields, new SW modes are observed in the continuous regions due to the non-synchronous rotation of the magnetization with respect to patterned areas[131]. Unconventional structures like magnonic quasicrystal in the form of octagonal lattice of antidots[132], binary ADL with alternating hole diameter[133] and defective honeycomb lattice[134] have been developed in the pursuit of greater tunability of SW spectra and anisotropy. Figure 5 shows the lower frequency SW modes of octagonal ADLs exhibit an unconventional eight-fold rotational anisotropy superposed with a weak two-fold and four-fold anisotropies. Significantly, the contribution of eight-fold anisotropy gradually reduces with the increase in lattice constant. New measurement techniques like all-electrical measurement by using inverse spin Hall effect (ISHE)[135] have also fuelled new interests in this field. Potential applications using ADL includes development of highly reprogrammable magnetic array[136] and tunable magnonic filter[137] integrated in a magnonic waveguide[138].

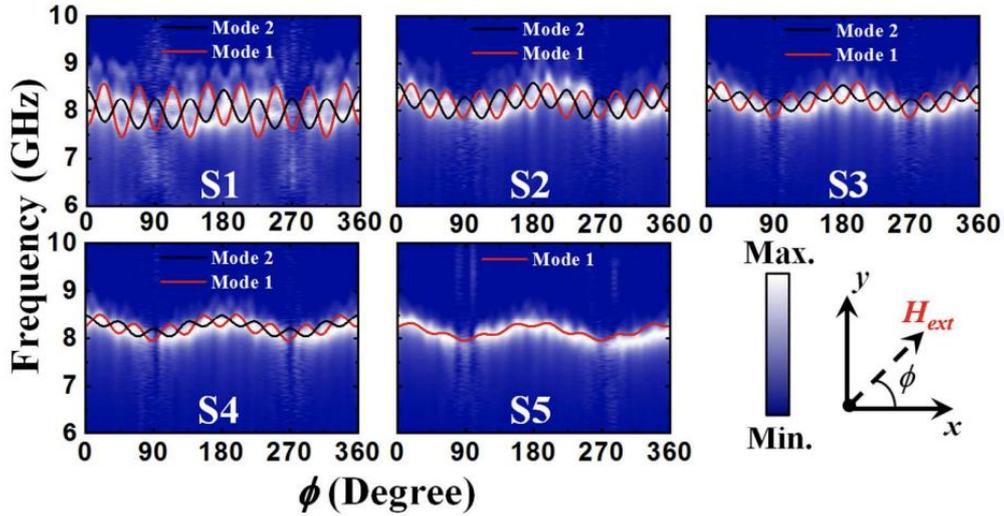

FIG. 5. Variation of spin wave (SW) frequency with the azimuthal angle ($\phi$) varying from 0° to 360° for NiFe ADLs with various lattice spacing (S1-S5) are presented at bias magnetic field, $H_{ext}$ = 800 Oe. The surface plots represent the experimental results while the solid lines describe the sinusoidal fits for the observed anisotropic SW modes in all the samples (S1-S5). The color map associated with the surface plots and the schematic of the orientation of the external applied field ($H_{ext}$) are given at the bottom right corner of the figure. Reproduced with permission from S. Choudhury *et al.*, ACS Nano **11**, 8814 (2017). Copyright 2018 American Chemical Society [132].

**Magnetic rings**

Ferromagnetic nanorings are high-symmetry structures in which magnetization forms a circulation state without the presence of a vortex core as opposed to a disk. The magnetic normal modes in nanorings are also simpler than those in disks. The study of spin dynamics in ring structure is also of direct practical importance, as the circulation direction in ring structures has been proposed to be used in vertical magnetic random access memory cells[139]. They can have two kinds of magnetic states: a flux closure or "vortex" state and an "onion" state with same moment orientation in each half of a ring and they exhibit a range of different switching mechanisms including DW and vortex core nucleation, annihilation and propagation etc. Initial studies showed that the transitions from onion-to-vortex and vortex-to-reverse onion states are strongly dependent on the edge-to-edge-spacing of the rings due to dipolar interaction as well as shape anisotropy[140]. The seminal work on the dynamics of ring structure showed the excitation of two dominant modes by uniform perpendicular and in-plane pulsed field, respectively a circularly symmetric uniform mode and a rotationally antisymmetric mode, while other modes are of very small amplitude[141]. In the same year Giesen *et al.* showed that depending on the ring width, a splitting of the uniform precession mode occurs, and the high-intensity FMR modes are localized in specific segments of the ring[142]. Investigation of spatially resolved dynamic eigenmode spectrum in Co rings revealed up to five resonant modes in the frequency range from 45 MHz to 20 GHz as a function of an external magnetic bias field. The vortex and onion states led to well-defined and distinctive mode structures, which were affected by the dynamic inter-ring coupling[143]. The applied magnetic field has also been used to split the radial and azimuthal excitations due to either mode localization or symmetry[144]. Further distinct series of quantized azimuthal modes in a vortex state stemmed from the constructive interference of circulating SWs were observed. This can be considered as a magnetic ring resonator resolved up to fourth order[145]. Observation of coherence and partial decoherence in Py rings in the onion state was mapped by microfocused BLS[146]. The shape of the ring added additional complexity in the static and dynamics and as opposed to a circular ring, triangular[147] and square[148] nanorings showed larger number of modes, the nature and frequencies of which strongly depend on the orientation of the applied magnetic field. A width-dependent transition from radial to azimuthal modes was observed in the square ring. Anti-rings are another type of interesting structure and direct mapping of static and dynamic magnetizations their potential application in biosensing have been reported[149]. More recently, SW mode conversion and mode hopping in anti-ring structures has been observed[150], leading towards their potential use in magnonics.

**Bi-component magnetic nanostructures**

Interestingly the first report on MC was a composite or bi-component medium where a ferromagnetic material embedded in a ferromagnetic background (Fe cylinders in an EuO matrix) showing MBG[151]. Although theoretical works progressed in this field[152] it was not until 2009, when MBG in a 1D bicomponent stripe (Co/Py) was revealed[153]. The size tunability of MBG in 1D BMC was an important development[68]. Soon after this, investigation on 2D BMC was started and Co nanodisks embedded in Py antidots showed two channels of SWs, one through the Co nanodisks and another in between them[154]. In a parallel work on SW dispersion in 2D BMC consisting of Co square dots embedded in Py matrix showed larger frequency width of magnonic bands than the constituent antidots and a complicated magnonic band structure, where the Co dots act as amplifiers of dipole coupling between the Py dots[155]. A major challenge in the fabrication of 2D BMC was to achieve direct physical contact between the two different magnetic materials. Choudhury *et al.*[156] achieved this feat in Py-filled CoFe ADLs, where they obtained strong signature of inter-element exchange interaction across the interface and ensuing enhancement of the SW propagation velocity by a factor of three as opposed to the ADL only. The shape of BMC has also been varied in this study and a more recent report on triangular BMC showed strong anisotropic SW[157]. In an important development, 2D CoFeB/Py bicomponent lattices have been used as an omnidirectional nanograting coupler, which shows an enhancement of the amplitude of the short-wavelength SWs as compared to a bare microwave antenna[25].

**Binary magnetic nanostructures**

The improved functionalities of BMCs come at the expense of more complicated fabrication processes, such as multistep lithography and TPL. An alternative structure, namely binary nanostructure can provide similar flexibility along with simpler single-step lithography process by placing two structures of same or different material next to each other forming the basis of the crystal. A Py/Co binary nanostructure grown by a simple self-aligned shadow deposition technique showed rich SW spectra and large anisotropy of SW[158]. A binary nanostructure in the form of diatomic nanodot array (dipolar-coupled Py nanodots forming a complex double-dot basis) showed excellent tunability of the magnonic band structure by changing the orientation of the bias magnetic field in the BLS measurements. New interaction modes appeared in this structure due to strong dipolar coupling between the inter-dot interaction in the diatomic unit[159]. A strong anisotropy in the SW dispersion is observed in this structure, which is also evident in the iso-frequency curves, promising their applications in magnon focusing and defocusing[160]. Observation of spectral narrowing and mode conversion in novel binary nanostructure has been reported where square-shaped Py nanodots of two different sizes are diagonally connected to form a binary basis[161].

**H. Dynamics of nanoscale magnetic multilayers**

Magnetic multilayers (MMLs) offer a huge versatility in quantum magnetic properties, including spin-dependent scattering, spin tunneling, exchange anisotropy, orbital hybridization, perpendicular magnetic anisotropy (PMA), interfacial Dzyaloshinskii-Moriya (iDMI) interaction, pure spin current, SHE, Rashba-Edelstein effect, spin caloric effect, voltage controlled magnetic anisotropy (VCMA), topological spin textures, etc. A variety of existing or proposed spintronic devices are based on MLs, namely giant magnetoresistive sensor, magnetoresistive random access memory (MRAM) cells, STNO, racetrack memory, and non-reciprocal magnonic devices. In all form of devices nanopatterning is the key. However, inclusion of this huge field in this sub-section is beyond our scope. Hence, we discuss three important aspects, namely patterned PMA structures, STNO structures and pure spin current driven nanomagnet dynamics using SHE.

MMLs with PMA are prospective candidates for bit patterned media. They also have applications in skyrmions, VCMA and various other fields. Although significant works on the magnetization dynamics of thin film MML with PMA have been reported[162-165], very few efforts have been made on the dynamics of nanostructured MML with PMA. Some resports exist on magnetization reversal and DW study with defects and edge corrugations[166], increase of coercive field[167] in Co/Pd antidots and DW pinning in Co/Pt antidots[168]. Theoretical study of SW in patterned MMLs with PMA using discrete dipole approximation showed that magnetic inhomogeneity along the central axis splits the magnetostatic SWs into two bands, and the exchange SWs into ae number of bands as a result of the underlying long- and short-range interactions, respectively[169]. The first experimental observation of SW dynamics in this system came in 2014, when Pal *et al.* observed a decrease in SW frequency with increasing density of antidots, down to values well below the FMR frequency of the continuous ML, in a series of [Co(0.75 nm)/Pd(0.9 nm)]$_8$ ADLs with PMA[164]. This was modelled by assuming nanoscale rim-like shell regions surrounding the antidots created by the Ga$^+$ ion bombardment during patterning using FIB technique. The decrease of SW frequencies is found to be driven by a dynamical coupling between the localized modes within the shells, most likely by tunnelling and exchange interactions. The shape of the antidots has found to play decisive roles in the in-plane domain structure and the ensuing edge-localized SW spectra, their mutual interactions and interaction with bulk SW excitations[170].

It is quite fascinating that spin-polarized current or pure spin current can apply torque on the magnetization in MML which can excite oscillatory magnetic modes in the nanomagnets. In a magnetic tunnel junction-based STNO this microwave oscillation is converted to voltage from the change in magnetoresistance. In SHNO, pure spin currents drive local regions of magnetic films and nanostructures into auto-oscillation. These microwave-generating nanomagnets can act as nanoscale motor, resonator, transducer etc. The first experimental demonstration was reported by exploiting a heterodyne mixer circuit for 130×170 nm$^2$ elliptical nanopillar made of Cu(80)/Co(40)/Cu(10)/Co(3)/Cu(2)/Pt(30) MLs[59]. The achievable power could be tuned by varying the current and magnetic field for a wide frequency range. Additionally, on tuning several factors the dynamics entered from linear to nonlinear regime. However, the microwave power emitted from single STNO was found to be only 1 nW. Thus, necessity for designing mutually phase locked array of oscillators became crucial. The mutually synchronized oscillators showed a sudden narrowing of the signal linewidth and an increase in the power due to coherence[23]. This caused noise reduction and increased stability analogous to the array of Josephson junction oscillators. The output power can scale with $N^2$ at room temperature where

$N$ is the number of oscillators. Thus, collective dynamical response from the array of nano oscillators can be extremely useful in powering a larger network and neuromorphic computing[171].

Successful generation of microwave in ML nanomagnets can be achieved in another energy-efficient manner. The antidamping-like torque exerts a negative damping on the magnetization and can eventually overcomes the intrinsic damping showing auto-oscillation around the effective magnetic field. However, suppression of the auxiliary modes showing nonlinear trends must be deliberately avoided. Demidov. *et al.* demonstrated construction of such SHNO in specially designed ML structures and explained the intriguing mechanism for a single nanomagnet[22]. The same group developed a nano-notch SHE oscillator directly incorporated into a magnonic nano-waveguide for simultaneous excitation and enhancement of propagating SW in the waveguide[172]. It has further been demonstrated that by controlling the PMA strength, one can suppress the nonlinear magnetic damping[10], for achieving decay-free propagation and probable amplification of SWs[173].

**I. Dynamics of nanoscale spin textures**

Spin textures are nonuniform spin configurations in magnetic material, which are stable, resilient and yet possess remarkable degree of tunability and scalability. They are extremely promising for energy-efficient, dynamically reconfigurable and reprogrammable components in spintronics and magnonics. Here we briefly discuss the progress in magnetization dynamics of spin textures, namely DW, vortex and skyrmion.

**Domain wall**

Magnetic DW is a common magnetic texture with intriguing physical properties, which has attracted huge scientific interest due to its potential application in magnetic logic devices and topology-based memory applications[174]. Motion of the DW in response to high frequency has drawn attention since 1950s[175]. Later, current induced dynamics and switching of domains attracted attention, and domains in adjacent cobalt layers were manipulated controllably between stable parallel and antiparallel configurations by applying spin-polarized current pulses of the appropriate sign[176]. Imaging of DW oscillation by TRSKM showed two different oscillation modes at 0.8 and 1.8 GHz to be concentrated at different parts of the DWs in a microscale magnetic element[177]. Subsequently, intrinsic nonlinearity in the resonant response of magnetostatically coupled transverse DWs was demonstrated[178].

Interaction of SW with DW has also become an important topic. Hertel *et al.* numerically showed that the presence of a 360° DW causes strong attenuation of the radiating SWs due to pronounced phase-lag between the propagating wavefronts along the two arms of a nanoring[179]. Similar phase-shift was also predicted for the SWs colliding with 180° Bloch-DW[180]. Kim *et al*. reported that monochromatic dipole-exchange SW undergoes a peculiar negative refraction due to collision with the 90° DW at the twin interface of 5-nm-thick Fe film with cubic anisotropy[181]. Using LTEM imaging, Sandweg *et al*. demonstrated modulation of thermal SWs in presence of asymmetric transverse DW[182]. Further BLS measurements revealed annihilation of quantized SW modes near the DW and evolution of new mode inside the complex SW structure due to the change in the effective internal field within the DW region[183]. The focus gradually shifted towards the investigation of DW-assisted propagating SWs to construct reconfigurable magnonic nanocircuitry. Garcia-Sanchez *et al*. reported a remarkable observation of guiding of propagating excitations localized to the wall in curved geometries while flowing in close proximity to other channels[184]. For Néel-type walls they show strong non-reciprocity and an analogy with the whispering gallery modes of sound wave. Woo *et al*. illustrated DWs as stationary reservoir of exchange energy, that can be generated, manipulated, and used to release on-demand SWs[185]. This can further be detected by using DWs in an all-DW device. Recent emergence of application of DWs as reconfigurable MC and SW nanochannels has been described in section III. I.

**Magnetic vortex**

In systems with negligible magneto-crystalline anisotropy, the competition between demagnetization and exchange energy may lead to the flux closure structure like magnetic vortex with in-plane curling magnetization and out-of-plane core within ferromagnetic planar element with certain dimensions (see Fig. 6 (a))[45]. The core polarity ($P = \pm1$) and in-plane circulation (clockwise or counterclockwise) can be manipulated by magnetic field, current or spin torque to make these topological solitons fundamentally interesting and viable for applications[186,187]. Initial study of time domain measurement of vortex dynamics in Py square elements with lateral dimension down to 500 nm revealed a high frequency mode originating from the uniform precession of the magnetization, which is associated with another low frequency mode characteristic of the spiral gyrotropic motion of the vortex core due to magnus force[177]. Rigid vortex model considering the edge poles and another model avoiding the edge poles, were used to describe these motions. BLS measurements and hybrid micromagnetic modelling revealed the presence of azimuthal SW modes in magnetic vortex states in Py dots with 100 nm diameter and 15 nm thickness[95]. Further experiment revealed splitting of the azimuthal modes into doublet due to the coupling between SWs and the gyrotropic motion of the vortex core[189]. Dynamics of vortex with a pinning site showed gyration of the vortex about a single pinning site at low excitation amplitude, gyration due to the magnetostatic energy of the entire vortex for high excitation amplitude and a sharp transition between these two amplitude regimes that is due to depinning of the vortex core from a local defect[190]. In 2010, a frequency controlled magnetic vortex memory was proposed[191].

Gyration of coupled magnetic vortices turned out to be an exciting problem and was first reported by Buchanan et al. where two vortices were found to stabilize in a single elliptical element. Four eigenmodes were observed, which were identified as the in-phase and out-of-phase polarization-dependent core gyration[192]. A flurry of papers on coupled magnetic vortices came after few years using different techniques[193], which showed that the coupled pair of vortices behave like a diatomic molecule with bonding and antibonding states, promising the possibility for designing the magnonic band structure in an array of magnetic vortex oscillators[194,195]. Barman *et al.* numerically showed efficient energy transfer of gyration mode in a locally excited 1D chain of physically separated nanodisks[196] which was later experimentally verified by Hasegawa *et al.*[197]. Jung *et al.* experimentally demonstrated logic operations based on this idea[198]. In 2014, a novel magnetic vortex-based transistor (MVT) operation was proposed by Kumar *et al*. There, a three-vortex sequence (material Py: diameter = 200 nm; thickness = 40 nm; separation = 50 nm) with polarity combinations (1, −1, −1) was considered as an MVT, providing a gain (*i.e.* the amplification of core gyration amplitude from input to output vortex) of ~15 dB similar to an electronic bipolar junction transistor (BJT)[199]. The gain was further optimized by tuning the intervortex separation in asymmetric three-vortex (AMVT) network[200]. Furthermore, the output from this AMVT was fed into the input of another two AMVTs to perform a successful fan-out operation and nearly equal gains were achieved. Exploiting the asymmetric nature of the energy transfer mechanism by stray-field antivortex solitons, a successful fan-in operation with construction of a tri-state buffer gate was recently reported[201]. Figure 7 shows the simulated magnetostatic stray field distribution of a single vortex, coupled vortices, AMVT, AMVT-based networks arranged in fan-out and fan-in configurations. Collective modes in 3D magnonic vortex crystal have also been observed recently[202].

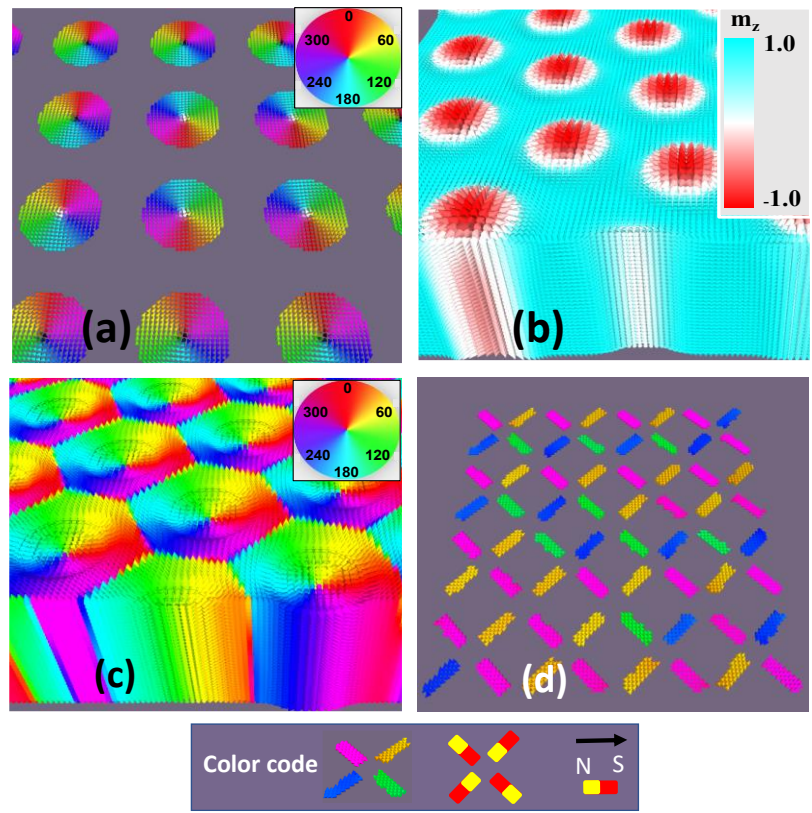

FIG. 6. (a) Simulated spin configuration of 2D array of magnetic vortices with different polarity and circulation combination. Simulated skyrmion lattice: (b) in-plane (x-y) angle variation of magnetization shows a hexagonal lattice symmetry. (c) distribution of $m_z$ component shows formation of skyrmion tubes along its thickness. Color maps are presented inside the figures. The simulation parameters are taken from C. Wang *et al.*, Nano Lett. **17**, 2921 (2017) [188]. (d) Simulated spin configurations of a 2D artificial spin ice (ASI) structure at remnant state. The color code is indicated at the bottom of the figure.

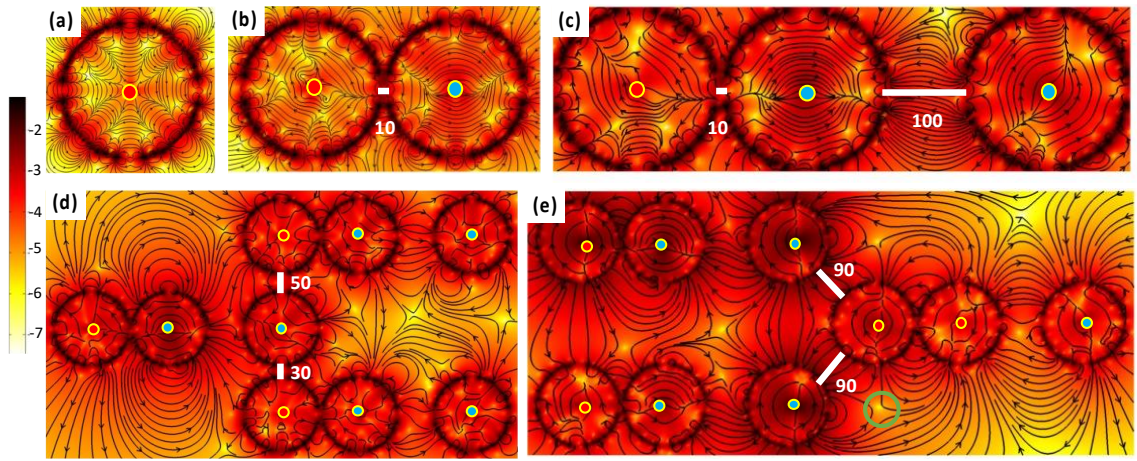

FIG. 7. Magnetic field distributions for (a) single vortex (diameter: 200 nm and thickness: 40 nm), (b) magnetostatically coupled two vortex systems with inter-disk separation = 10 nm, (c) AMVT structure with inter-disk separations 10 nm and 100 nm, (d) fan-out unit and (e) fan-in unit. The images are obtained after 25 ns of the excitation provided at the input vortices. The excitation field of amplitude 1.5 mT is applied for all the systems. One antivortex in the stray field is marked with green circle for representation. The color bar is shown at the left of the figure. The contour coloring is based on the sum of squares of x and y components of the magnetic field and the color bar is in dB. The up and down polarities of the vortices are marked by red and blue dots, respectively. The important physical distances are indicated in numeric figures where the unit is nm.

**Magnetic skyrmions**

Skyrmion is a topologically protected swirling magnetic texture with a topological charge or winding number ($S = \frac{1}{4\pi} \int \boldsymbol{m} \cdot (\partial_x \boldsymbol{m} \times \partial_y \boldsymbol{m}) dx dy = \pm 1$) that emerges within the bulk non-centrosymmetric compound or nanoscale MMLs with broken inversion symmetry at the interface. These are found to be very robust under external perturbations. DMI plays a pivotal role in stabilizing the skyrmions. They are of two different types: Bloch and Neel skyrmions. After the first theoretical prediction back in 2001, skyrmions were first observed in bulk non-centrosymmetric MnSi crystal in 2009[203] and Fe monolayers and PdFe bilayers on Ir (111) in 2011[204]. The exploitation of spin-polarized current or pure spin current have been found to be extremely energy efficient way to trigger the dynamics of these solitons. The fact that skyrmions can exhibit fast longitudinal motion at an average speed of 100 m/s plays a key role in designing of skyrmion-based racetrack memory and many other magnetic devices.

Theoretical calculations showed various types of dynamical modes of skyrmions, namely clockwise, counterclockwise and breathing mode and a melting of skyrmion crystal when intensely excited[205]. Using microwave transmittance spectroscopy, a nonreciprocal directional dichroism effect via skyrmion resonance modes in a helimagnetic $Cu_2OSeO_3$ was observed[206]. Further, broadband ferromagnetic resonance study of collective spin excitations (magnetic helix and skyrmion dynamics) in the metallic, semiconducting and insulating chiral magnets and a precise quantitative modelling across the entire magnetic phase diagrams of the systems quantified the chiral and the critical field energy[207]. Büttner *et al*. used time-resolved pump-probe X-ray holography to image gigahertz gyrotropic eigenmode dynamics of a single magnetic bubble whose trajectory confirmed its skyrmion topology. The trajectory further revealed a large inertial mass of the skyrmion due to its topological confinement and the energy associated with its size change[208]. Time-resolved pump-probe soft X-ray imaging technique revealed distinct dynamic excitation states of 100-nm diameter magnetic skyrmions, triggered by current-induced SOT. The dynamics of magnetic skyrmions was found to be efficiently controlled by the SOT on the nanosecond time scale[209]. Current driven dynamics of frustrated skyrmions in a synthetic antiferromagnetic bilayer showed interesting results. While the in-plane current-driven bilayer skyrmion moves in a straight path, the out-of-plane current-driven bilayer skyrmion moves in a circular path with better in-plane current-driven mobility of a bilayer skyrmion than the monolayer one[210]. Very recently exciting development of the coherent propagation of spin excitations over a distance exceeding 50 µm along skyrmion strings in the chiral-lattice magnet $Cu_2OSeO_3$ has been reported. The propagation is found to be directionally non-reciprocal and the degree of non-reciprocity, group velocity and decay length are strongly dependent on the character of the excitation modes[211].

## J. Dynamics of artificial spin ice structures

Artificial spin ice (ASI) are engineered materials where ferromagnetic nano-islands are arranged in a predefined geometry to manifest frustration. In such systems, there is an ambiguity to choose the magnetic ground state, because of degeneracy which gives rise to the 'zero-point entropy'[212]. The frustration leads to the creation of magnetic monopole defects connected through 'Dirac string' with the anti-monopole defects. The natural monopoles were first observed in pyrochlore material ($Ho_2Ti_2O_7$)[213] resembling the frustrated geometry of water ice, but those were difficult to control externally. The first report on ASI system in 2006[214] revealed that experimentally the monopole states can be stabilized in a square ASI system. This was followed by a series of studies including square and kagome ASI lattices as can be found in the literature[215,216]. Static characterizations showed that defects in ASI are very much sensitive to the magnetic history and applied magnetic field which can influence the SW dynamics. The SW dynamics with changing monopole numbers and string length in square ASI was first analytically reported by S. Giga *et al.*[217]. The monopoles modify the frequency of EMs and splitting of the mode signifies monopole resonance. Later on, researchers observed these dynamics experimentally[218-220]. Field dependent hysteresis explored the complex behavior of low frequency mode at lower bias magnetic field in square ASI[218]. Two different SW modes were found to coexist at zero field depending on the magnitude of the initial applied field. The occurrence of monopoles manifested new SW mode in an interconnected kagome ASI system[219]. The SW mode intensity increased with increasing number of monopoles. A systematic study of static and dynamic properties of six different spin ice and anti-spin ice was reported by Zhou *et al.*[220], where the sensitivity of SW modes to the aspect ratio in ASI systems was observed[221,222]. Mode merging was observed with decreasing thickness of nano-islands in square ASI[221]. The different shape anisotropies of nano-islands in Y-shaped units of the kagome lattice[222] were found to control SW modes' activation and deactivation depending on the applied magnetic field direction. The structural parameters in ASI clearly influence the wave-vector dispersion. A theoretical calculation of topology-controlled SW dispersion in square ASI was reported by Iacocca *et al.*[223]. The experimental demonstration of SW dispersion in ASI was reported in the literature.[224,225]. The weak intra-element coupling showed flatter dispersion[224]. The opening of channel at 45° angle led to dispersive SW nature in a square anti-ASI system[225]. The extensive study of SW dispersion in the strongly coupled and interconnected ASI may open up new opportunities for constructing ASI-based magnonic devices. To realize the magnetic frustration in 3D, square spin ice geometry with vertically shifted nano-islands[226], quasi tetrahedral geometry of 3D inverse opal-like structures[227], diamond bond-like 3D lattice mimicking the ASI structure[228] have been developed. However, the SW dynamics in 3D ASI systems is yet to be realized.

**K. Dynamics of 3D magnetic nanostructures**

3D nanomagnetism is a new avenue for future magnetism research. The inclusion of new dimension in 3D magnetic nanostructure (3D MNS) gives rise to different novel phenomena such as Bloch point[229], curvature induced anisotropy[230], magnetic monopoles and its charge conservation in 3D[231], topological structures[232], magnonic band structure[233,234] and others. On the other hand, 3D MNSs are the potential candidates for ultrahigh data storage capacity and processing devices[235], building blocks for neuromorphic computer architecture[236], magnetic sensors and actuators[237] etc. The main hindrances on the path of 3D nanomagnetism are the fabrication and characterization of 3D MNSs. To this end, some of the 2D fabrication techniques have been promoted to fabricate 3D MNS. TPL is a powerful tool to fabricate scaffold of 3D nanostructure. By combining TPL with various deposition techniques, complex 3D MNSs have been fabricated[33,228]. The main issue with this technique is the accessibility of the full geometry. Focused Electron Beam Induced Deposition (FEBID), commonly known as 3D-nanoprinting is a versatile tool for 3D nanostructure fabrication. Any arbitrary shape can be fabricated with FEBID technique. The main issue with FEBID is that the purity of the deposited material reduces due to contamination with carbon and oxygen. High quality nanowires were fabricated using FEBID and MOKE signal from them was measured [238]. Chemical procedure has also been used for 3D MNS fabrication[239]. Here, the main constraint is that the sample is deposited on a prepatterned template. 3D tetrapod magnetic structure made with four connected wires, has been studied by the X-ray[240] and electron[241] tomography techniques to map the magnetic induction and magnetization vector field of 3D MNS, respectively. Element specific mapping ability of magnetic circular dichroism with X-ray[242] and electron beam[243] have also been used to study 3D MNSs. Some recent literatures[26,244] have made detailed review of 3D MNS. The magnetization dynamics of 3D MNS is a necessity for understanding of SW characteristics and their applicability in spintronic and magnonic devices. Periodically patterned 3D nanostructured can form 3D MC. Theoretical study by Mamica *et al.*[233] revealed that the magnonic band structure in 3D MCs is very much sensitive to saturation magnetization, Heisenberg exchange stiffness constant and anisotropy. An alternative way of realizing 3D MC is in the form of patterned MLs, the SW dynamics of which have been discussed already in section 2H. Sahoo *et al.* studied the magnetization dynamics from the junction of a Co tetrapod structure using TRMOKE microscope (see Fig. 8)[245]. With the help of micromagnetic simulations they explained the nature of different SW modes appeared in the experimental SW spectra. However, the experimental study of collective magnetization dynamics of periodically arranged 3D MNS is still missing in literature. Broad aspects of 3D magnonics have been discussed in ref. 246.

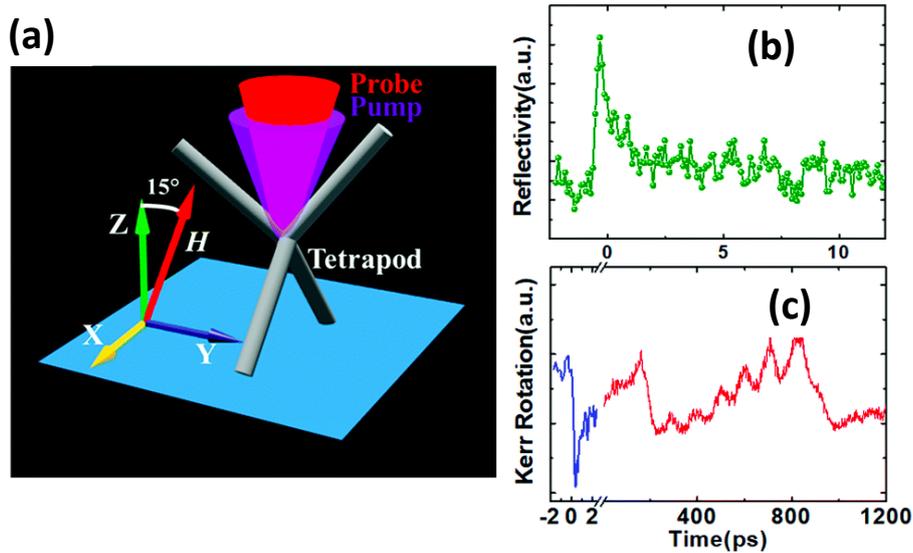

FIG. 8. (a) Schematic diagram of the cobalt tetrapod and the experimental geometry. Typical time-resolved (b) reflectivity and (c) Kerr rotation data are shown at $H$ = 3.92 kOe. Reproduced with permission from S. Sahoo *et al*. Nanoscale, **10**, 9981-9986 (2018) [246].

### III. OUTLOOK OF EMERGING PHENOMENA AND FUTURE CHALLENGES

The last one decade had witnessed a profusion of exciting developments in the field of nanomagnetism as discussed in the previous sections. Nonetheless those bring along new challenges and hurdles along the path of exploring novel and intricate physical phenomena as well as elegant and demanding applications starting from imaging the ultrafast dynamics of a tiny magnet to the development of an all-magnetic computer. In this section we will discuss about some of these emerging phenomena and challenges in the field of magnetization dynamics of magnetic nanostructures.

**A. Spatiotemporal imaging of magnetization dynamics of magnetic nanostructures**

The continuous development of spatiotemporal imaging techniques has made it possible to study femto and picosecond dynamics of single nanomagnet well below the diffraction limit. However, spatial mapping of the confined modes of individual nanomagnets, collective precessional modes in arrays, short wavelength magnons, dynamics of spin textures like vortex and skyrmions, SW caustics, bullets, ultrafast spin accumulation in nanomagnets, etc. need further development of both benchtop techniques in individual laboratories and large facility-based techniques. Various probes have been used for these measurements, which include electrical, optical, X-ray, force and electron microscopies. The electrical techniques are easy to use and commercially viable, but lack both the required spatial and temporal resolution. Development of spatially resolved FMR[247] raised some hope to this end due to the exceptional versatility of the FMR techniques but the progress have been very slow beyond that[248]. Time-resolved MOKE offers unprecedented temporal resolution and with the advent of attosecond laser the temporal resolution can get even better. BLS technique has its advantage of measurement of thermal magnons and even when external excitation is used, no synchronization is needed between the optical probe and the external source. Improvement of numerical aperture of microscope objective such as solid immersion lens[249] and lowering the wavelength of the probe laser could take the spatial resolution down to about 200 nm. The development of scanning near field optical microscopy (SNOM) and its integration with magneto-optical measurements[250] ushered hope to this end. However, imaging of localized edge mode in ferromagnetic element by near-field BLS microscopy[251] have drawn criticism[252]. Moreover, fiber and aperture based SNOM suffers from depolarization effect and lack of photon efficiency, as the the optical transmission of apertures with sub-wavelength diameter is proportional to $(r/\lambda)^4$ where $r$ and $\lambda$ are the radius of the aperture and the wavelength of the light, respectively. The role of localized surface plasmons in enhancement of near-field MOKE signal will be significant here. Further development of near-field AFM probe, commercial supply of smaller aperture, polarization preserving plasmonic antenna, such as bow-tie and cross bow-tie geometry, would be required for routine measurements with sub-100 nm spatial resolution[39]. The experimental detection of intensity and phase maps of SWs for specific SW modes in nanoscale elements are still open challenges. The techniques those require large scale facilities, such as, electron or neutron scattering with superior spatial resolution cannot be implemented on tabletop. Space-, time- and phase-resolved BLS can be very powerful in this aspect but require improvement of resolution[253]. An important problem is to measure the wave-vector dispersion of SW frequency from nanoscale elements. However, in BLS microscopy uses microscope objective with high numerical aperture, where it is non-trivial to resolve wave-vector information. However, using Mach-Zehnder interferometry the phase information can be retrieved.

Electron microscopy is a powerful technique, which can provide atomic resolution. Addition of time resolution gives rise to ultrafast electron microscopy (UEM) or dynamic transmission electron microscopy, (DTEM) which can be the strongest tool for the study of ultrafast dynamics in nanomaterials[254]. The use of photoactivated electron source driven by a nanosecond laser gives rise to few nanosecond time resolution, which can potentially be improved to picosecond and further to femtosecond. Time-resolved two-photon photoemission spectroscopy[255] is a powerful tool developed to study electron dynamics in metal and has been used to study nanostructured silver film, for example[256]. It has the potential for application in magnetic nanostructures with suitable modifications. Development of polarization determination of light emitted from nano objects by means of polarized cathodoluminescence (CL) spectroscopy with an ellipsoidal mirror in a transmission electron microscope[257] would be important for measurement of time-resolved magneto-optical signals using time-resolved electron microscopy.

Polarized X-ray microscopy has revolutionized magnetic imaging. The magnetic contrast from X-rays comes from XMCD for ferromagnetic material and XMLD in case of antiferromagnetic materials. The focusing of X-rays are generally done by Fresnel zone plate (FZP) as well as multilayered Laue lenses, refractive lenses, or zone-doubled diffractive optics. The two most popular FZP-based techniques, namely magnetic full field transmission soft X-ray microscope (MTXM)[258] and a scanning transmission soft X-ray microscope (STXM)[259] are being widely used to study magnetic DW and vortex dynamics, while their applications to skyrmion dynamics and SW propagation have just been started. Further advantages like element specificity, interfacial sensitivity and 3D tomographic capabilities with nm resolution can give new directions in nanomagnetism and spintronics which are fast moving into 3D arrangements of spin structures, non-collinear spin arrangements and SOTs in ultrathin film heterostructures and MLs. Resonant soft X-ray ptychography offers to retrieve real space image with 10-nm resolution[260]. In all these techniques successful integration with femtosecond laser will lead to unprecedented spatiotemporal resolution. Soft X-ray laser with a high harmonic generation has been proven to be a viable route for strongly enhanced coherent extreme ultraviolet radiation with femtosecond pulse duration associated with a photon flux of $3\times10^{11}$ at 32.8 nm[261]. Such a high flux is sufficient for single shot imaging of nano objects.

Among other promising techniques SPSTM allows to image spin textures at nearly atomic resolution and it has recently been extended to the time domain to study fast electron spin relaxation times occurring in the ns regime[262]. Lensless imaging of magnetic nanostructure using X-ray spectro-holography is a form of Fourier transform holography, which is transferable to a wide variety of specimens, scalable to diffraction-limited resolution, and is well suited for ultrafast single-shot imaging with coherent X-ray free-electron laser sources[263]. MRFM bridged the gap between MRI and SPM[264]. It operates on the principle of probe-beam modulation reflected from a scanning tip with about 700 nm diameter sensing the magnetic field from the sample. On the other hand, nanomagnetic tip attached to MRFM facilitates scanning with sub-nm spatial resolution and in near future it can be employed for 3D imaging[265]. We hope that further development can lead this technique to be established as a reliable microscopy in FMR mode for studying dynamics of nanomagnets. Nitrogen vacancy (NV) centre is an atomic sized point defect in the diamond crystal providing high spatial resolution and also possessing remarkable magnetic and quantum properties with high-field sensitivity and excellent spatial coherence. NV centre magnetometry[266] is another powerful nano-magnetometric tool due to its above properties and can be useful for imaging nanoscale spin dynamics, if coupled with high frequency.

**B. Arrays of nanostructures with complex geometry**:

Like all other crystals, MC also enjoys the benefits of introducing geometrical complexity to achieve extremely rich and complex SW dynamics and magnonic band structures. This includes the introduction of varying lattice symmetry as well as complex basis structures. When added to the inherent magnetic anisotropy, non-uniform internal magnetic field, interfacial effects etc., these can lead to almost unlimited possibilities for MCs, which make them unique among its counterparts from other artificial crystals. In section 2G, existing reports on SW dynamics in various forms of complex 2D arrays of nanostructures such as octagonal lattice, defective honeycomb lattice, binary nanostructures, rings, anti-rings etc. have been discussed. Composite antidot structures with continuously varying lattice constant[119,129] have also been studied scarcely. Although some attempts have been made to vary the lattice symmetry as well as the basis structures[267,268], that form only negligible fraction of the overall possible combination of such structures and phenomena. Therefore, systematic variation of various geometric parameters for tuning the SW dynamics and magnonic band structures will be extensively studied in the coming years too. An example is given in Fig. 9, where triangular antidots are arranged in honeycomb lattice with varying lattice constant. The combination of an asymmetric basis and lower lattice symmetry shows a magnonic spectra which shows excellent tunability with lattice constant. Along this line if we introduce binary or ternary basis with antidots with different size, shape or separation, we can further complicate the SW spectra and mode structures. They can introduce new bands or (semi)localized states in the magnonic band structure, besides modulating the MBG very efficiently. In photonics less symmetric structure has been found to give better self-collimation[269] and slight modulation of lattice symmetry or basis structure can give rise to new phenomena useful in different applications. The simulated equilibrium magnetic configurations, SW spectra and SW power maps of two exemplary complex basis nanostructures are shown in Fig. 10.

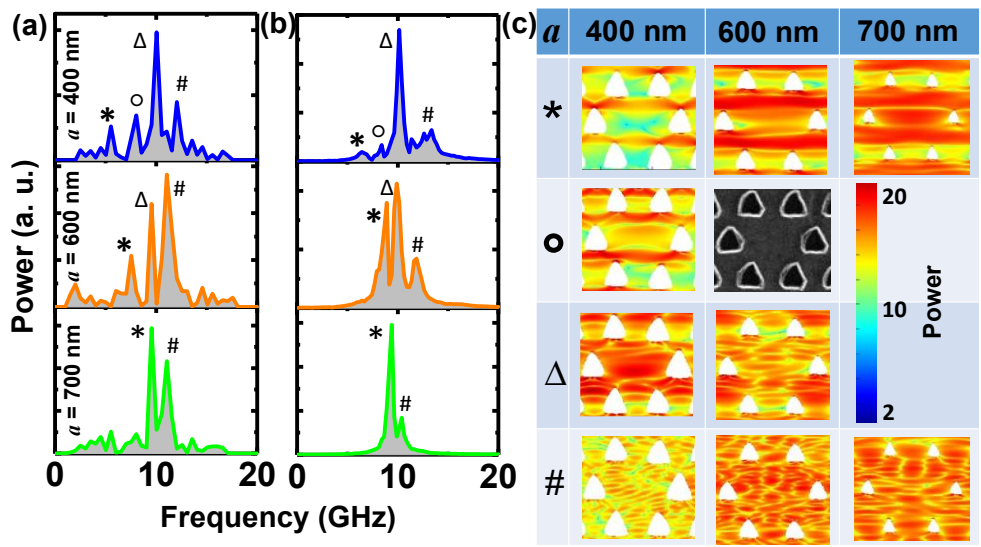

FIG. 9. (a) Experimental and (b) simulated power spectra of a triangular shaped antidot lattice having honeycomb symmetry with varying lattice constants ($a$). (c) Simulated power profiles of the observed SW modes. The color map used for the power profiles and the SEM image of the lattice with $a = 400$ nm are shown inside the figure. The number of modes and the overall bandwidth of the spectra reduces with the increase in lattice constant. The lateral width and power of the extended mode through the channels between the antidots increase and the quantized mode also become more prominent with the increase in $a$.

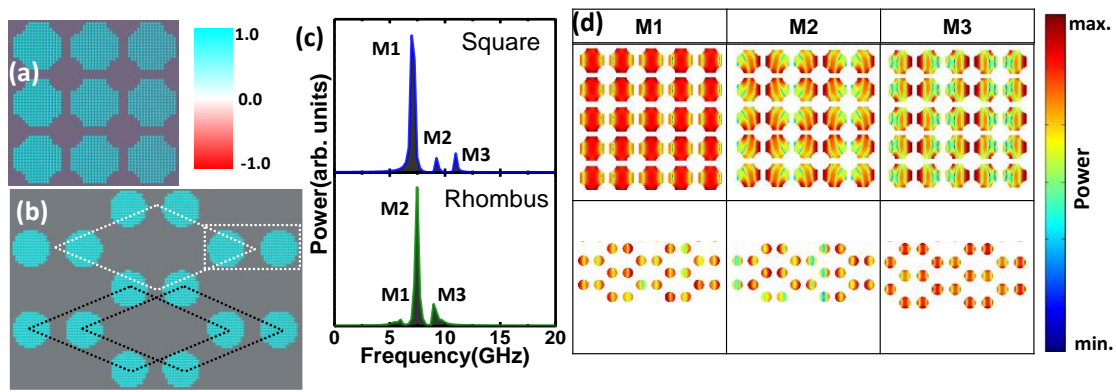

FIG. 10. Simulated static magnetization configurations of two different complex basis nanomagnet arrays at a bias magnetic field 1 kOe: (a) square lattice with complex basis of length 125 nm and lattice constant 150 nm and (b) two-atom basis rhombus lattice with dot radius of 100 nm, interdot separation in the basis of 50 nm and lattice constant of 300 nm (highlighted by white dashed line) or superposition of two linearly shifted rhombus lattice (highlighted by black dashed line). (c) Simulated SW spectra for square and rhombus lattices at a bias magnetic field 1 kOe. (d) The corresponding simulated SW power profiles are shown. The color bars are presented besides respective figures.

The SW mode profiles show some complex characters that are not observed in usual basis structures as described in section 2G. Further complexity can be added to the 2D arrays by inclusion of spatial nonuniformity of the magnetic properties within the individual nanomagnet. This can be termed as 'Janus nanomagnet' analogous to Janus nanoparticles having symmetric shape but asymmetric surface properties[270] (see Fig. 11 (a)). Such structural engineering may require complicated fabrication processes such as, multistep lithography. However, the powerful exchange coupling within the nanomagnet itself made of two different materials can introduce stark variation in the SW spectra and will be worth exploring in near future.

**C. Arrays of nanostructures with quasiperiodicity and defects**

Quasicrystals possess long-range ordering without any periodicity and their diffraction patterns exhibit symmetry forbidden by crystallographic restrictions. Quasicrystals have been extensively studied in photonics and phononics for a long time but have only recently been introduced in nanomagnetism in the form of magnonic quasicrystal (MQC). Appearance of pass band[271], allowed bulk band in place of band gaps[272], modulation of magnonic gaps[273], damping of collective SW modes[274] etc. have been theoretically investigated in 1-D bi-component MQCs with Fibonacci sequences. Grisin *et al*. fabricated Fibonacci type structures consisting of grooves and the crests made in YIG[275]. Apart from multiple forbidden gaps, the appearance of relatively narrow pass bands for the MSSW mode in their measurements led to the development of a ring resonator. Bhat *et al*. studied MQCs made of Py interconnected nanobars arranged in Penrose P2, P3 and Ammann tiling, which exhibited distinct sets of FMR modes with characteristic angular dependencies (eight- and ten-fold rotational magnetic symmetry) for applied in-plane magnetic fields[276]. Choudhury *et al*. observed characteristic magnonic spectra and an eight-fold rotational symmetry in a Py antidot lattice with octagonal symmetry[132]. Lisiecki *et al*. observed a remarkable dynamic coupling between propagating SWs through Py nanowires of two different widths arranged in a 1D Fibonacci sequence using STXM measurement[277]. This field, however, remained wide open with new opportunities. For example, a crystallographic structure constrained by only two simple physical properties, 'discreteness' and 'homogeneity' turns into Delone sets[278]. Magnonic analogue of these sets can be easily realized to explore intriguing SW dynamics in this unconventional structure. Another interesting polytope of MC can be 'Voronoi cell'-like nanostructures, which retains the lattice symmetry but offers more independent choice of the basis[278]. Figure 11 shows the ground state spin configurations of some of these structures. Construction of numerous types of bi-component or multi-component MQCs is possible with aperiodic geometries, such as, different variants of penrose tiling, oblique tiling and Ammann-Beenker tiling[278,279] which may offer unprecedented tunability of the magnonic band structure besides stark modulation of SW group velocity due to the lack of translational symmetry.

MCs with 'fractal' geometries can now be included as a new member of the family of artificial crystals offering intriguing high-frequency dynamics. In 1919, German topologist, F. Hausdroff introduced the idea of fractional dimension, and in 1975, Mandrot assigned the name 'fractal' to a particular class of curves[278]. Those are the curves whose paths are monodimensional, however at the limit they occupy a 2D area. Dai *et al.* have recently reported micromagnetic simulation study on magnetization reversal and magnetic spectra in Sierpinski triangles[280]. However, experimental results are still lacking in this field. An exemplary Sierpinski's closed peano curve in the form of a MC is shown in Fig. 12. The simulated SW spectra spreads over a broad range between 5 and 17 GHz for moderate value of bias magnetic field. The highest frequency mode exhibits mixed azimuthal quantization within the elements and a 'Fractal' like spatial distribution over the whole array. Such structures are expected to provide enormously rich SW spectra with broad range of control parameters and exotic SW propagation properties.

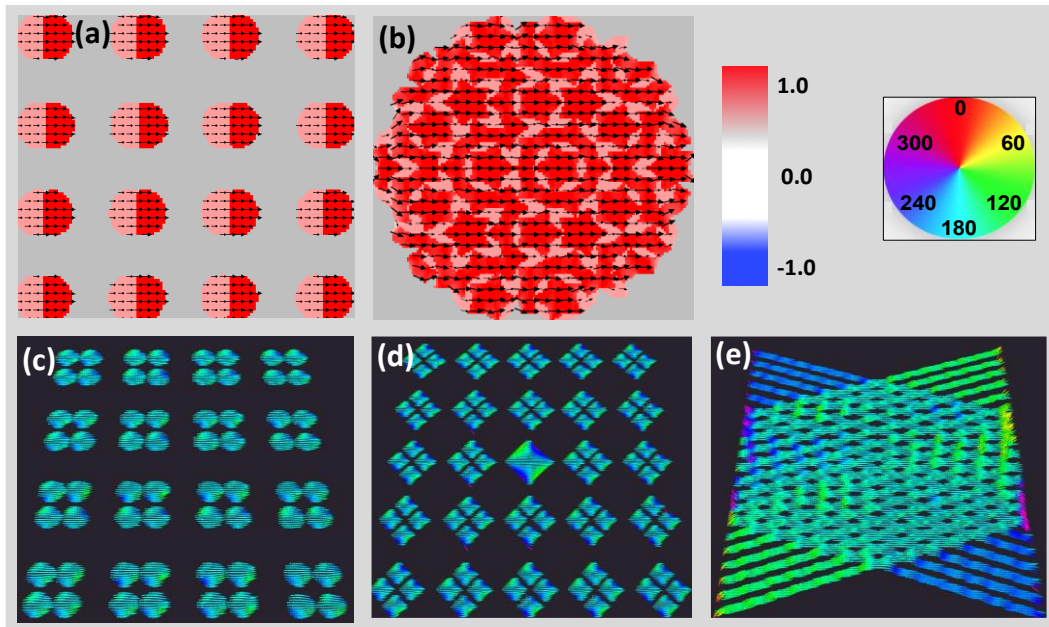

FIG. 11. Ground state spin configurations for (a) bi-component MC with Janus nanomagnets as basis, (b) bi-component MQC with Penrose tiling arrangement, (c) MC resembling symmetrical Delone set, (d) 2D array of diamond shaped nanodots with point defect, (e) cross-weaved nanostripes: structural defect arising due to overlapping of nanostripes with moderate tilting. Color bars are presented at the top-right corner of the figure.

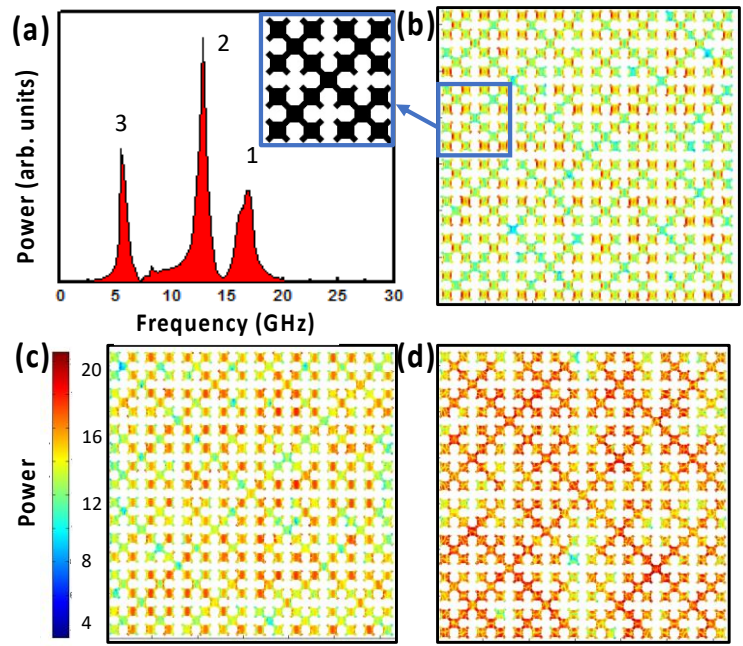

**FIG. 12.** SW dynamics of Sierpinski's closed peano curve is simulated in form of a MC with the material as Py and array size of 3×3 µm² discretized into 5×5×20 nm³ cells. (a) Frequency spectra for 'fractal' structure at bias magnetic field of 1 kOe. Mode numbers are indicated in numeric figures. A portion of the structure is shown at the inset. SW power profiles are shown for (b) mode 3, (c) mode 2, and (d) mode 1. The corresponding color bar is shown at the left of the figure.

The tunability of magnonic bands in MCs is the key for efficient transfer and process of information in magnonics. Introduction or intrinsic occurrence of defects or disorder reforms the local magnetic properties including magnetic potential. This may drastically reconstruct the magnonic band structure by introducing new SW modes, phase shifts, minibands, and adjusting MBG and SW group velocities due to the breaking of translational symmetry. Kruglyak *et al*. numerically showed that introduction of an isolated defect layer in a 1D MC leads to the appearance of several localized defect modes within the band gaps[281]. Ding *et al*. experimentally studied the correlation between FMR response (singlet or doublet) and the degree of disorder in a 1D array of dipolar-coupled Py nanowire[282]. Such defect modes are often found to be dispersionless[283]. Point defects with symmetrical shapes and moderate filling fraction embedded within 2D MC have generated multiple defects modes[284]. It can be expected that the defect with inferior or no lines of symmetry (such as, rectangle, triangle, diamond, cross, parallelogram etc.) and superior filling fraction may offer greater tunability of magnonic bands. Morozova *et al*. showed that length of magnetization localization region depends on line-defect width, which can be used for developing SW logic and multiplexing blocks[285]. The edge roughness and shape deformation occurred due to the limitation of nanofabrication primarily affects the edge modes and localized SW modes[131]. Magnonic spectra of Co antidot lattice with hexagonal symmetry was found to be quite robust to introduction of random defect[129]. Further case simulations on square antidots (100 nm width) arrange on square lattice (lattice constant = 200 nm) showed a sharp variation in SW spectra with the emergence and subsequent splitting of defect modes confined within the defect regions (see Fig. 13). Reduction of antidot size and lattice constant, when the SW transforms into dipole-exchange mode, caused a significant variation in the pattern of defect density dependence, where only slight blue-shift in an extended mode and gradual disappearance of quantized mode occur. Defects in MC can play significant role in magnonic bands analogous to electronic bands in semiconducting crystals and extensive research to this end is expected in the coming years.

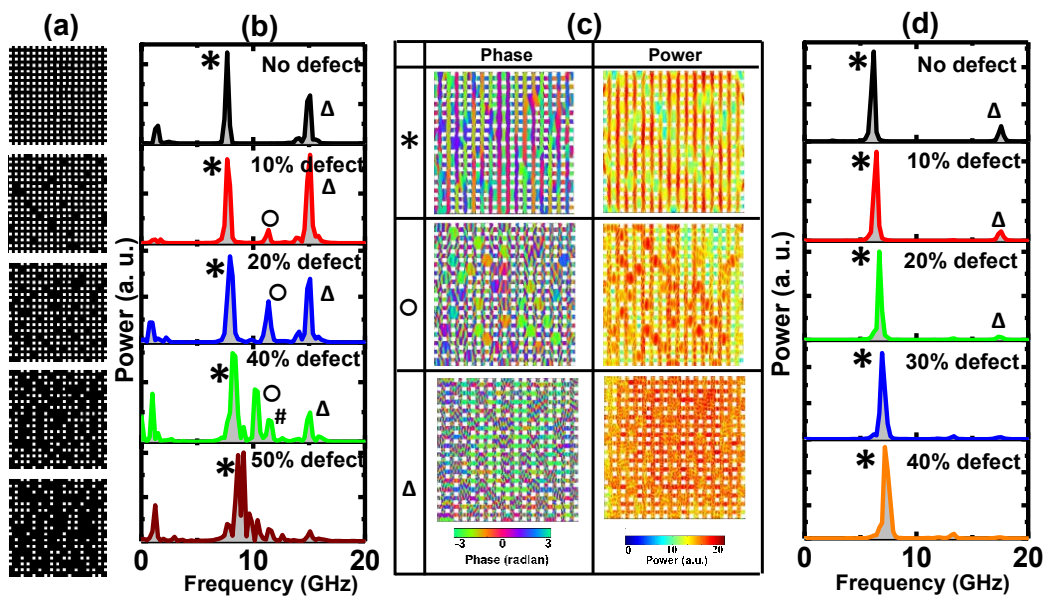

FIG. 13: (a) Images of ADLs with defects having hole diameter of 100 nm and separation of 100 nm. (b) Simulated SW spectra of the corresponding ADLs with defects. (c) Simulated SW power and phase maps of the same ADL with 20% defect. (d) Simulated SW spectra of the ADLs with defects having hole diameter of 25 nm and separation of 50 nm.

**D. Spin-orbit coupling in nanoscale spin dynamics**

SOC effect has emerged as a very rich and elegant effect in nanomagnetism and spintronics[286]. It enhances significantly with reduced dimensions due to broken inversion symmetry at the surface or interface to produce spin-split dispersion (Rashba SOC), chiral spin texture (iDMI), spin-polarized surface states with topological properties (topological insulators), spin-momentum locking among many others. Various other intrinsic and extrinsic SOC based properties such as SHE, spin pumping, PMA, etc. exist which have made strong inroads into present or future technologies based on spin-charge conversion, stabilization of skyrmions, STNO, SHNO etc. Many of these strongly affect the spin dynamics starting from ultrafast demagnetization, relaxation phenomena, SW propagation and damping. Hence, it is imperative to extend the application of single or multiple of these properties to our advantage in excitation, control and detection of the dynamics of nanoscale spin textures and patterned nanomagnet arrays. For example, simultaneous existence of PMA and DMI in patterned 2D nanomagnet array can help in stabilizing topological magnetic objects like skyrmions, bubbles or merons and more efficient control of their dynamics. Development of 1D and 2D MCs on ferromagnet/nonmagnet heterostructures along with iDMI will lead to novel magnonic band structures and evolution of new bands due to the asymmetric SW dispersion in presence of iDMI (see Fig. 14). Pure spin current driven and controlled coherent SW is nanomagnet arrays would also be an engrossing research area.

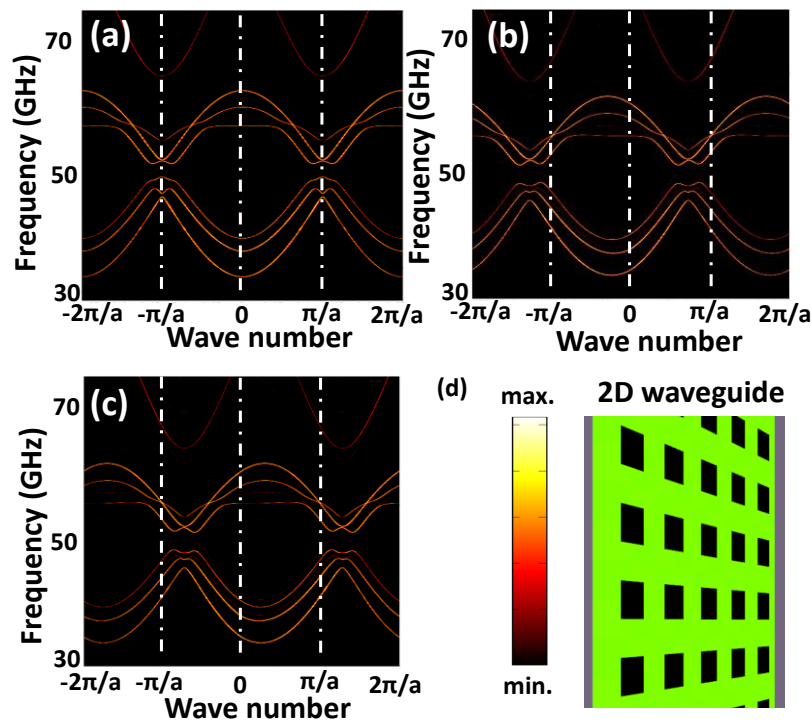

FIG. 14. Simulated SW dispersion curves of a quasi-2D antidot waveguide in the DE geometry in presence of bias magnetic field = 1.01 T with DMI ($D$) values of (a) $D = 0$ mJ/m$^2$, (b) $D = 1$ mJ/m$^2$ and (c) $D = -1$ mJ/m$^2$. (d) A schematic of antidot waveguide and color bar of SW amplitude are presented at the bottom right corner. Simulation parameters are, waveguide dimension: 12 μm × 120 nm × 3 nm, cell size: 6 nm × 6 nm × 3 nm, AD hole size: 12 nm, lattice constant: 24 nm, $M_s$: 800 kA/m, $A_{ex}$: 13 pJ/m.

Topological insulators[287] with extraordinary properties like ambipolar surface states, spin-momentum locking, protection from backscattering between states of opposite momenta with opposite spins and ensuing reduced energy dissipation, high spin-charge conversion will lead to exotic properties and devices such as low power electronics like topological p-n junctions[288], SOT-MRAM, nano-batteries, topologically protected skyrmion states, abacus-type applications, etc. Further magnetic doping or proximity of topological insulators may break time-reversal symmetry, leading to a gap at the Dirac point and a reorientation of the low-energy spin texture and investigation of quantum AHE[289] would be exciting prospects in this field. In graphene, SOC can be enhanced due to proximity and hybridization with an adjacent magnetic layer both due to intrinsic Rashba SOC and extrinsic defect mediated SOC. In heterostructure of graphene with another 2D material such as transition metal dichalcogenides with larger intrinsic SOC[290] or a topological insulator[291], the prospects of spin-charge conversion associated with larger mobility, conductivity and longer spin lifetime of graphene would be very exciting.

Development of a hardware network which can be successfully trained by a set of STNOs to recognize spoken letters by tuning the frequency and amplitude according to the real-time learning rule with excellent scalability[236,292] promises the construction of future artificial neural networks. A practical STNO-array based device will probably require current from a single source distributed to each oscillator through parallel or series connections. These devices will find applications as sources in nanoscale phase-locked arrays[293], which could be used in wireless chip-to-chip or intra-chip communications. SHNOs can be more advantageous due to easy integration on Si substrate and the exploitation of pure spin current with better energy efficiency.

**E. Emerging phenomena in quantum hybrid systems**

Hybrid systems have rapidly emerged as strong candidates for quantum information processing[294] where quantum states are coherently transferred between two media using different carriers such as superconducting qubits, spin ensembles, optical and microwave photons and phonons. For example, microwave photons in high-quality cavities are very efficient for communicating spin information because of their long coherence length, whereas other quasiparticles such as magnon and phonons can also generate indirect interactions between tiny magnets over long distances. In the following, we discuss the recent progress and future directions in this field.

**Magnon-phonon coupling**

The coupling between magnonic and phononic degrees of freedom in radio frequency regime was theoretically predicated long back[295] but it has attained great momentum more recently. Over the last decade, SAW in GHz regime have been implanted to excite or manipulate the

SW in magnetic thin films[296,297] and nanostructures[62,64,298] through magneto-elastic (ME) interaction. SAW driven ferromagnetic resonance has been studied in ferromagnetic film[296,297] and single nanomagnet[299,300]. The SW frequency[64], damping[301] and field dispersion[298] can be tuned by ME coupling. SAW-modified rich SW texture has been observed in a magnetostrictive single nanomagnet[64]. Strong coupling between magnonic and phononic degrees of freedom has been observed in a single nanomagnet[62]. SAW-driven dynamic MC generation based on Bragg scattering and Doppler shift has been demonstrated[302]. On the other hand, elastically driven spin pumping[303], phonon mediated inverse Edelstein effect[304], the mechanism of phonon-magnon coupling such as spin-rotation coupling[305] and role of time-reversal symmetry[306] have stirred huge interest in this field. Magnon-phonon coupling plays a key role in the spin Seebeck effect, where modification of magnon distribution function by phonons results in a pure spin-current injection from a ferromagnet to nonmagnet[307]. In many of these phenomena, size reduction and geometric parameters will play key roles in exploring new and superior functionalities. The condensation of mixed magnon-phonon state and magnon-phonon bottleneck accumulation phenomenon[308] are extremely rich and fundamental phenomena. Moreover, the potential of its occurrence in any multicomponent gas mixture of interacting quasiparticles having different scattering amplitudes promises the exploration of new systems in near future.

**Magnon-photon coupling**

Magnon-photon coupling or cavity magnonics has versatile applications such as hybrid quantum systems[309], coherent conversion between microwave and optical frequencies[310], quantum electrodynamics[311] and direct detection of dark matter[312]. In 2014, Tabuchi *et al.* observed hybridization between ferromagnetic magnon and microwave photon in the quantum limit[311] and thus leading the way to magnon-photon coupling. In 2015, Viennot *et al.* coherently hybridized the individual spin and charge states of a double quantum dot while preserving spin coherence up to the megahertz range at the single-spin level in a superconducting resonator[313]. To overcome the issue of limited lifetime of magnon-polariton, Zhang *et al.* created stationary magnon-polariton states by a dynamical balance between pumping and losses with non-Hermitian spectral degeneracies[314]. However, these works are predominantly based on YIG, which is non-trivial to integrate on chip and suffered from lack of proximity between magnetic material and microwave resonator. Li *et al.* overcame this problem by fabricating an all-on-chip magnon-photon hybrid circuit with a Py thin film device directly fabricated on top of a coplanar superconductor circuit[315] and obtained coupling strength of 0.152 GHz and a cooperativity of 68. In another back-to-back article Hou *et al.* used lithographically defined superconducting resonators to demonstrate high cooperativity between a resonator mode and Kittel mode in a Py wire with number of spins ~$10^{13}$. This was soon followed by another article by Wang *et al.* who showed cooperative effect of coherent and dissipative magnon-photon couplings in an open cavity magnonic system. Their system leads to nonreciprocity, flexible controllability as well as unidirectional invisibility for microwave propagation[316]. Overall, the field of cavity magnonics is heating up but several problems remained open. Two most important issues are downscaling of the magnetic system (number of spins) while retaining high enough cooperativity and identification and quantification of coherent and dissipative coupling origin, strength and their competition, which will surely receive intense interest in next few years.

**Magnon-magnon coupling**

Strong magnon-magnon coupling has set-off very recently and still in its nascent stage. Initial study of excitation of exchange-coupled perpendicular standing spin wave (PSSW) in YIG from the FMR frequency of an adjacent Co layer by using strong coupling and avoided crossing has been demonstrated in 2018[317]. Similar avoided crossing, a signature of strong coupling, was observed between the PSSW modes of YIG film and FMR mode CoFeB film[318] and that of Ni nanowires[319] at the same time. The latter work is important as it is a way forward to overcome the issue of 'coupling strength' which can be enhanced by square root of number of spins ($N$) to overcome the weaker coupling strength ($g_0$) between individual spins and the microwave field, i.e. $g = g_0\sqrt{N}$[311]. Subsequently, avoided crossing for interlayer coupled atoms in two-dimensional antiferromagnet $CrCl_3$[320] have been demonstrated. More recently, coherent spin pumping and large avoided crossing in YIG/Py bilayer with reduced layer thickness[321] and magnon-magnon coupling in interlayer exchange coupled synthetic antiferromagnets of FeCoB/Ru/FeCoB layers[322] have been achieved. However, strong coupling between nanoscale metallic ferromagnetic elements is essential for on-chip integration of hybrid systems. To this end a breakthrough was obtained very recently, when a very strong magnon-magnon coupling between Py nanocross elements at moderate microwave power has been achieved where two anticrossings between magnon modes have been observed[110]. Further numerical simulations demonstrated strong coupling between EM and CM in single nanomagnet[323]. The last two works are expected to open pandora's box in the dynamics of nanomagnetic systems where intra- and inter-element interactions can be tailored by various material and geometric parameters of nanomagnets to engineer the strong magnon-magnon coupling.

**F. Emergence of nanomagnet antenna**

Antennas act as an essential component in smart phones, tablets, biologically implanted devices, radio frequency identification systems, radars, etc. They are an array of conductors that can generate the oscillating electric field and magnetic field through oscillating electric currents which are required to ensure a high radiation altitude. Conventional antennas rely on electromagnetic wave resonance that leads to antenna sizes comparable to the wavelength λ. Further miniaturization of antenna size has been one of the fundamental challenges for years, as antennas substantially smaller than the wavelength of electromagnetic wave exhibit decreased radiation resistance, large electrical Q-factors, and decreased radiation efficiency due to additional losses such as Ohmic dissipation[324,325]. These limitations have made it extremely challenging to achieve miniaturized compact antennas at very-high frequencies (VHF, 30–300 MHz) and ultra-high frequencies (UHF, 0.3–3 GHz), putting severe obstacles on the wireless communication systems. Thus, new antenna concepts need to be investigated with mechanisms for the miniaturization of antenna size.

An alternative avenue to reduce antenna size is to excite an electromagnetic antenna at acoustic resonance instead of electromagnetic resonance. Strong strain-mediated magnetoelectric coupling in ferromagnetic/piezoelectric heterostructures enables efficient energy transfer between magnetism and electricity[326]. This concept has recently been adopted by Nan *et al.* to design acoustically actuated nanomechanical magnetoelectric antennas with such ferromagnetic/piezoelectric thin film heterostructure, which receive and transmit electromagnetic waves through the magnetoelectric effect at their acoustic resonance frequencies[327]. Since the acoustic wave velocity is roughly five orders of magnitude smaller than the velocity of light, consequently, the acoustic wavelength is five orders of magnitude smaller than the electromagnetic wavelength at the same frequency. Hence, the radiation efficiency ($A/\lambda^2$, where $A$ is the emitting area and $\lambda$ is the emitted wavelength) of such antennas are several orders of magnitude larger than the electromagnetic antenna, which allows drastic miniaturization of communication systems. A few reports on antenna miniaturization mechanisms using this concept have been demonstrated[328,329]. Another antenna miniaturization concept using unique fractal geometry has been proposed[330]. More recently, an antenna implemented with closely packed array of magnetostrictive nanomagnets deposited on a piezoelectric substrate has been developed[331], where a SAW launched on the substrate with an alternating electrical voltage can periodically strain the nanomagnets and rotate their magnetizations owing to the Villari effect (Fig. 15 (a)). The oscillating magnetizations of the nanomagnets emit electromagnetic waves at the frequency of the applied SAW. Clearly, such miniaturized antennas can drastically enhance antenna gain at small size, allow dramatic downscaling and are expected to have great impacts on future communication systems. However, tuning the geometrical and structural parameters and excitation methods can play important roles in exploring better functionalities of such nanomagnet antennas. Thorough experimental and theoretical investigations are thus highly demanded in future to overcome several open challenges of miniaturized nanomagnet antennas with better prospects.

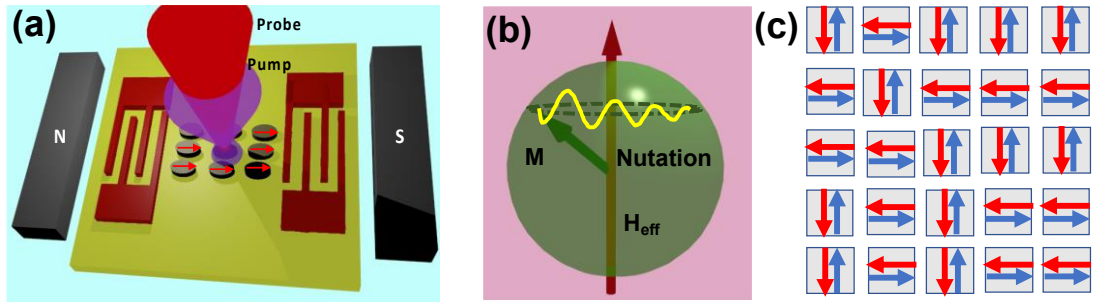

FIG. 15. (a) Schematic of optical pump-probe technique with nanomagnet array fabricated in between two Interdigitated switches in presence of magnetic poles. (b) Schematic of nutation dynamics of a nanomagnet. (c) Array of antiferromagnetic nanomagnets.

### G. Nutation dynamics in nanomagnets

Magnetization dynamics has similarity with gyroscope dynamics. Hence it is expected to observe a high-frequency motion along with the precessional dynamics in ferromagnetic nanostructures commonly known as 'nutation'. The familiar classical analogue is a spinning top. If an external field suddenly transfers angular momentum the moments get diverted from the angular-momentum axis and start to nutate on a fs time scale followed by the ps precession [Fig. 15 (b)]. This is a consequence of Heisenberg exchange interaction. The former effect is non-trivial to detect experimentally for having shorter amplitude and time scale. It is worth mentioning that nutation dynamics has been reported in various other systems, such as, superconducting Josephson junction[332], matter-wave vortices[333] etc. However, most of the available reports are based on the theory and simulations including those in nanomagnetic systems. Thus, this field is wide open offering scopes for extensive experimental research. Bottcher *et al*. demonstrated this intriguing physics theoretically for individual magnetic moment, a chain of Fe atoms and Co islands on Cu (111)[334]. There is a tradeoff between the nutation effect and inertia of the systems determined by an additional term in the atomic LLG equation. The nutation effect is very significant on the fs time scale with a typical damping constant of 0.01 to 0.1 in a single magnetic moment. In a chain of Fe atoms, the amplitude of the nutation depends on the number of interacting neighbours and with increasing damping both lifetime and magnitude of nutation decreases.

Numerical simulations by Olive *et al*. showed scaling function with respect to $\alpha\gamma H$ for the nutation angular frequency, which is also valid for the precession angular frequency when $\alpha\gamma H \gg 1$. Here $\tau$ is inertial dynamics characteristic time, $\alpha$ is the dimensionless damping, $\gamma$ is the gyromagnetic ration and $H$ is the static magnetic field[335]. This trembling of spins induced by inertial dynamics has been theoretically studied by various other groups[336,337]. A later report by atomistic approach revealed that the nutation dynamics can be induced by the nonuniform local fields originated from surface anisotropy in ferromagnetic nanoparticles[338]. Very recent experimental report revealed that a single YIG nanodisk (diameter: 700 nm) subjected to deep nonlinearity can accommodate stable quantized modes which do not transfer energy within themselves and thus are very energy efficient[264]. The spatial confinement of the nanomagnet helps to avoid any additional instability. A two-tone spectroscopic measurement was conducted where a strong continuous excitation along with a weak microwave field pulse modulated at the frequency of a cantilever tip, were used in the MRFM technique to probe the nutation on top of the steady state precession.

The experiments are still at a nascent stage and the scope is very vast, where nutation dynamics can be induced in single nanomagnets of varying magnetic potential as well as arrays of interacting nanomagnets of different geometric architectures. The excitation fields can vary from pulsed magnetic field, spin torque to acoustic excitations and very complex nutation consisting of multiple frequencies can be potentially obtained. Competition between inertial damping and Gilbert damping along with SOT-induced antidamping can be very exciting to follow in nanomagnetic systems. Analytical and numerical methods also need to progress hand in hand to capture such complex and fundamental dynamical processes.

**H. Emergence of antiferromagnetic nanostructures**

The experimental study of SWs in antiferromagnet[339] as well as theory of antiferromagnetic resonance[340] and linear SW theory[341] are known since 1950s. However, it has received great impact in recent times in spintronics and magnonics due to its various advantages over its ferromagnetic and ferrimagnetic counterparts. They do not have macroscopic magnetization and hence no stray field but can interact with spin-polarized current[342] and give rise to STT and skyrmion textures[343]. Due to their strong sublattice exchange field they can show huge resonance frequency in the THz range [344] and they are readily observed even in semiconductors and hence easy to integrate in hybrid devices using both charge and spin degrees of freedom.

Due to these advantages it is natural to develop, and study patterned antiferromagnetic nanostructures both from fundamental interest and from device fabrication. However, this field is still in its infancy. The recent discovery of long-distance spin transport through antiferromagnetic insulating materials opens up huge opportunities for emerging spintronics applications[345]. We envisage versatile application and fundamental study in patterned antiferromagnetic (AFM) structures starting from magnetic tunnel junction elements and MC to THz emitters. Recent theoretical study proposes the development of AFM-MC by using two complementary methods for achieving control over the magnonic degrees of freedom: a) by controlling the anisotropy properties of the SW system and b) by exposing the AFM to a tailored magnetic field. Formation of a band-like structure of allowed and forbidden bands, which can be adjusted by tuning the parameters of the MC, raises hopes for designing new types of MC complimentary to ferromagnet-based MC[346]. This issue of avoiding spurious 'cross-talk' even with high packing density is schematically shown in Fig. 15(c). For an array of AFM nanomagnet, the 180° switching of spins will not result any change in the net magnetization, but 90° rotation (along y axis in Fig. 15 (c)) of the Neel vector within the individual nanoelements can be considered as '0' or '1'. If these types of arrays are designed, the packing density of bits can be much larger than the ferromagnetic systems. However, the writing and reading of information within compensated antiferromagnetic nanomagnets in time scales of ps or less, will be far more challenging. Researchers are currently developing various methods involving charge current and spin currents for deterministic control of the ultrafast switching dynamics in AFM systems[347,348]. However, no such attempt has been made for patterned array of AFM nanomagnets yet. Determination of damping of AFM system is another challenge. This field is clearly brimming with many unsolved problems which need to be addressed in the upcoming years.

**I. Future direction in dynamics of spin textures**

As mentioned earlier, nanoscale spin-textures will play an important role in future reconfigurable and reprogrammable magnonics and spintronic devices. To this end, controlled preparation of magnetic DWs, vortex, skyrmions, bubbles and other spin textures, interconversion between them, their dynamics and interaction of them with spin current and magnon will dominate the magnonics and spintronics research in future.

Magnetic domains and DWs have shown great promises in racetrack memory, logic devices, SHNOs and huge efforts have been made in studying the DW propagation, expansion, pinning and depinning by magnetic field, charge and spin current[349]. This field will continue to grow with the introduction of various SOC effects as described in section III.D. The interaction of SW with DW has been less studied[179,350], but the interest is shifting towards this more recently. Recently DW has been proposed as reconfigurable MC due to scattering of surface SW at corkscrew type DW of aligned stripe domains in a Co/Pd ML with PMA[351] and SW nanochannels in a 180° Néel wall[352] and antiparallel coupled domain[353] arrangements. These areas will continue to grow with the introduction of various ML systems with PMA and synthetic antiferromagnets[354,355], where competition between Zeeman, PMA, demagnetizing and AFM energies and external perturbations can lead to easy tunability of domain structures.

Magnetic vortex dynamics in single vortex and coupled vortices have led to many interesting fundamental physics and application proposals as discussed in II.I. Research in this area has gained further momentum more recently with the interesting reports of whispering gallery magnons in vortex[356], chaos in incommensurate states of spin torque-driven vortex oscillations, MVT[199], etc. Fundamental studies on interaction of vortex with magnon[357], spin current, acoustic waves will continue to increase in the coming years. Numerical studies of coupled vortices demonstrated energy transfer, amplification, fan-in, fan-out, tri-state buffer switch operation and energy transfer by antivortex solitons[199-201]. Experimental realization of these effects by electrical and optical techniques will be important for further progress in this field. Moreover, dispersion of gyration modes and SW in 1D, 2D and 3D arrays of magnetic vortex (magnonic vortex crystal) will be a topic of intense interest. Local control of the corepolarization and circulation of individual vortex in an array may lead to huge set patterned vortex crystals. Spin-torque vortex oscillator [358] will be another field of interest and research on stabilization and dynamics of phase-locked vortex oscillators is expected to advance significantly.

Despite having a sizable effort, skyrmion dynamics and related effects will continue to emerge in future. A strong motivation in skyrmion dynamics is to achieve large skyrmion velocity in a magnetic track without appreciable drift along the transverse direction during the motion. Observation of clockwise, anticlockwise and breathing modes in single skyrmion of different size and skyrmion crystals of varying geometry are expected to receive much attention. Furthermore, hybridization of skyrmion modes with different magnon modes will also be imperative. The skew scattering of magnons from skyrmions gives rise to the topological magnon Hall effect (MHE), while topological magnus force leads to the skyrmion Hall effect (SkHE). A combination of MHE and SkHE may lead to a competition between thermally-driven radial magnon current, transverse magnon current and electron flow. By designing a magnetic track with in-plane anisotropy without DMI at the edges and out-of-plane anisotropy at the middle, SkHE can be reduced[359]. SkHE can also be suppressed in antiferromagnetically exchange-coupled skyrmions due to the absence of net topological charge and hence extensive research needs to be carried out with ferrimgnetic and AFM nanostrcutures. The annihilation and creation of skyrmions on application of electric field has also been proposed. Room-temperature devices such as, magnetic racetrack memory, skyrmion MCs, skyrmionic logic devices, skyrmion-based radiofrequency devices are few of those examples which will lead this field to its pinnacle in next few years[360]. Numerical calculation reveals that a chain of skyrmions exhibit rich SW band structure with their frequency ranging from 50-100 GHz[361], which is much higher than the conventional ferromagnetic resonance. Field controlled skyrmion crystals may act as reconfigurable and yet stable MC. Inertial mass originates from the ability of a system to store energy internally during its motion, and due to its additional topological source of inertia skyrmions can show very large inertial mass[208]. Observation of characteristic eignemodes of inertia and determination of inertial mass of skyrmions in various systems would be an important fundamental problem.

It is relevant to mention here that not only skyrmion but also half-skyrmion, antiskyrmion, skyrmonium, meron and other topological solitons are gaining huge attention[362-364]. For example, recent discovery of meron lattice in chiral magnet $Co_8Zn_9Mn_3$[365] raises the immediate challenge of creation and stabilization of a single meron pair, which is the most fundamental topological structure in any 2D meron systems, in a continuous FM film, and to study its dynamics. Finally, stabilization of skyrmions in unconventional systems like ferromagnetic thin films adjacent to 2D materials, topological insulators or heavy metals with weaker iDMI and excitation and control of their dynamics would be open problems.

**J. Progress with artificial spin ice**

The existence emergent magnetic monopoles in ASI systems is one of fascinating phenomena in magnetism. The back and forth movement of monopoles through Dirac string can be very beneficial for 'magnetronic' devices. So far, the monopoles have been mainly observed in connected and disconnected ASI. However, investigation of static and dynamic magnetic properties of a large variation of ASI structures both in 2D and 3D are foreseen in near future [215]. For SW propagation, connected ASI can be useful but the possibility of trapped DWs or even nonuniform spin textures at the junctions can pose hindrances to the propagating SW. The interaction of propagating SW with the monopoles will be an exciting phenomenon to explore. In addition to tunability of magnon band structure, possible scattering of magnons by magnetic monopoles may give rise to new type of topological Hall effect. In presence of additional stimulation, such as magneto-elasticity, spin torque, chiral spin texture, and other spin-orbit effects, the monopoles can lead to exotic properties of SW propagation. The ASI structures can also act as complex MCs. The magnetization configuration of ASI structures are highly sensitive to external applied field, leading towards exceptional tunability of SWs in such systems and formation of a new class of reprogrammable magnonic devices. The exploration of frustration in 3D [215,228] is promising both from fundamental physics as well as for extremely versatile tunable 3D MCs that can be useful for magnetic devices harnessing the variation of spin textures along the third dimension as well.

**K. Advancement in the dynamics of 3D magnetic nanostructures**

Despite its initial progress the fabrication of high-quality and versatile 3D MNS in sub-100 nanometer scale is yet to be achieved [26,228]. The FEBID is powerful technique to fabricate 3D complex nanostructure in the nanoscale, but the purity of magnetic materials needs substantial improvement. The combination TPL with electrodeposition or thermal evaporation can improve the material purity in 3D MNS but hitting the sub-hundred nanometer scale remained elusive so far. Modification of existing fabrication techniques and development of new techniques are pre-requisites for fabrication of high-quality 3D MNS in deep nanometer regime for the fabrication of compact 3D MNS-based magnetic devices. The ground state of 3D MNS may possess fascinating spin configurations such as hopfions[232]. The X-ray tomography[366] is capable of mapping the magnetization vector in 3D. The phase contrast imaging is possible with coherent X-ray[367]. The measurements in these high-end instruments depend on the availability of the techniques. New methods are required for faster measurements and laboratory-based characterization of 3D MNS. The SW in 3D MNS can be measured using benchtop techniques as discussed in sections 2C, 2K and 3A. Most of the available techniques are capable of measuring the SW dynamics from the surface of 3D MNS. To study the static and dynamic magnetic properties from the complex and angled 3D MNS, new techniques such as diffraction-based MOKE in off-specular geometry[368] or dark-filed MOKE magnetometry based on specular reflection[369] need to be further developed and integrated with time-resolved magnetometry. The experimental measurements of collective SW dynamics and magnon band structure in arrays of 3D MNS are still absent in the literature. New fabrication, experiment and analysis tools are pre-requisites to realize fascinating spin textures and SW dynamics in 3D MNS and development of practical devices based on them.

An exemplary numerical study of SW dynamics in a periodic 3D diamond bond lattice (DBL) constituted of square cross-sectional nanowires with 10 nm width and 25 nm length has been simulated by using mumax[3370]. Simulated spin configurations of different types of 3D magnetic nanostructures are presented in Fig. 16 (a), (b). The unit cell of the DBL is shown in Fig. 16(d), which consist of four tetrapods, each of which is made of four wires as shown in the inset of Fig. 16(d). Each tetrapod has a common junction where four nanowires are connected. An array of 3×3×3 DBL is shown Fig. 16(e). Each unit cell[371] has been assigned different colors for clear viewing. The simulated SW spectrum is shown in Fig. 16(f), which unraveled four clear SW modes. The phase profile of the highest intensity mode is shown Fig. 16(g). Within the individual nanowire, spins precess in the same phase, while the spins precess out-of-phase between the adjacent nanowires having an overlap at the junctions. Such complex spin dynamics need to be studied further experimentally and numerically to gain further insights in this burgeoning research field.

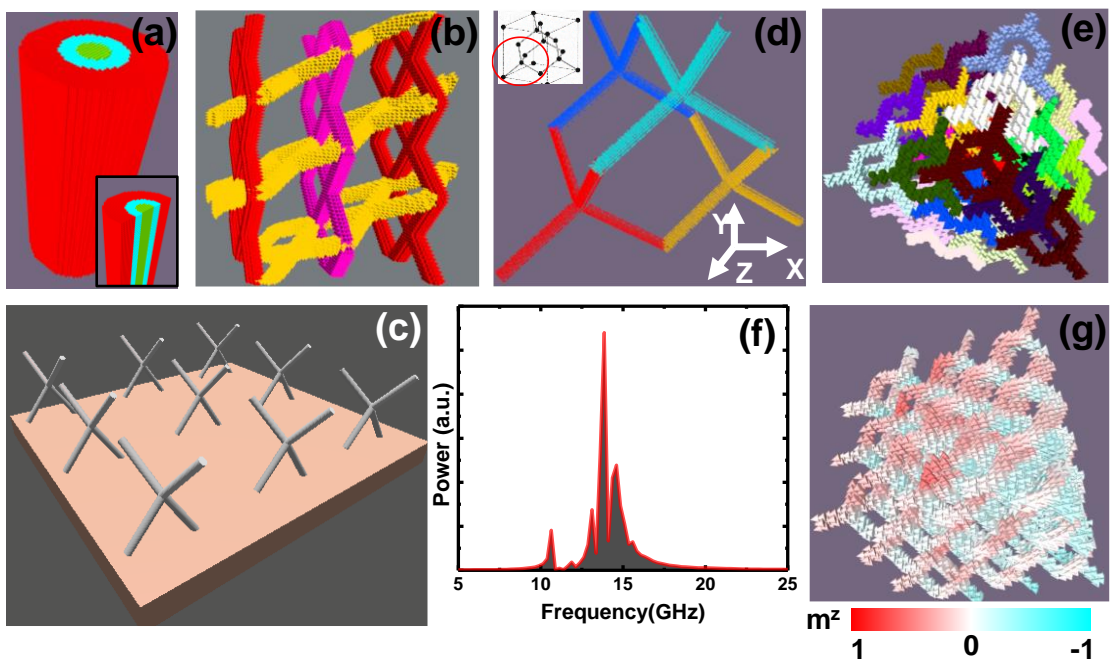

FIG. 16: Simulated 3D magnetic nanostructures. (a) Core-shell cylindrical nanowire with three different layers and (b) interconnected zigzag nanowires. (c) Schematic of 3D tetrapod structures arranged in a square lattice symmetry. (d) Simulated unit cell of diamond bond lattice. Different color is assigned to four different units for better visualization. Unit cell of diamond lattice is shown in the inset (adapted from Ref. 371), where one unit is highlighted by red circle. (e) 3×3×3 array of diamond bond lattice is shown, where each cube is assigned to different color for better visualization. (f) FFT power spectrum of simulated time resolved magnetization at an applied bias field of 4 kOe along x-direction. (g) The phase profile ($m_z$ component) of highest intense mode is shown.

# IV. CONCLUSION

Magnetization dynamics of magnetic nanostrcutures have been studied extensively for several decades now and during last two decades nanomagnetism has now grown remarkably to deliver diverse range of technological applications. Starting from the development of sophisticated fabrication processes and state-of-the-art characterization techniques to the construction of broad range of devices, the ceaseless progress in this field has spread its root deep into science and society. Achieving the feature size down to atomic scale limit and probing ultrafast phenomena in femtosecond time scale are no longer impossible. Besides the immense growth in 1D and 2D nanomagnetism, recent advancements and introduction of unconventional architectures, 3D magnetic nanostructures, magnon-spintronics, spin-orbitronics, topological spin textures in magnetic nanostructures have widen our horizon. To this end, the focus has recently been drawn more towards entangling multiple subfields together, which may lead to intriguing hybrid phenomena. Voltage controlled magnetic anisotropy, combination of spin transport with magnon dynamics, strong coupling in nanostructures, spin-texture driven spin waves are few of those examples.

Despite showing substantial promises, this field still offers an abundance of new potentials and faces stern challenges in reaching some of the overriding goals, like all-magnetic computing, neuromorphic and quantum computing, on-chip communication, etc. Integration of the magnetic components on-chip within atomically small area and minimization of power consumption are still massive challenge. Furthermore, fundamental areas like imaging of spin dynamics in nanomagnets, shot-wavelength exchange spin waves, spin accumulation, nutation and inertial dynamics need in-depth exploration. A combined theoretical and experimental groundworks are necessary in this sector. With a steep upsurge of consumer electronics in the coming years there will be a formidable competition between the charge-based and spin-based devices. Only time will tell if nanomagnetism along with its exceptional scalability of length and time, extraordinary diversity and flexibility of properties and low-power consumption is ready to become a leader in the 'post-Moore era'.

**Acknowledgement:** The authors gratefully acknowledge S. N. Bose National Centre for Basic Sciences, Kolkata and Department of Science and Technology, India for funding. The authors are grateful to Prof. Yoshichika Otani, Dr. Bipul Kumar Mahato and Dr. Dheeraj Kumar for their technical support in preparing some of the unpublished figures.

**Data Availability Statement:** The data that supports the findings of this study are available within the article.